\documentstyle[12pt]{article}

\newcommand{\vs}[1]{\rule[- #1 mm]{0mm}{#1 mm}}
\newcommand{\beq}{\begin{equation}}
\newcommand{\eeq}{\end{equation}}
\newcommand{\beqn}{\begin{eqnarray}}
\newcommand{\eeqn}{\end{eqnarray}}

\newcommand{\bmu}{\mbox{\boldmath $\mu$}}

\newcommand{\sect}[1]{\setcounter{equation}{0}\section{#1}}

\begin{document}

\begin{titlepage}

\hfill OUTP-01-21P

\hfill  April 2000

\vs{10}

\begin{center}

{\LARGE {\bf On the Microscopic Spectra of the Massive Dirac Operator 
for Chiral Orthogonal and Chiral Symplectic Ensembles}}\\[1.5cm]

{\large \sc{
Frank Abild-Pedersen$^{(1)}$ and Graziano Vernizzi$^{(2)}$}}\\[.5cm]
{\em $(1)$ The Niels Bohr Institute\\
Blegdamsvej 17, DK-2100 Copenhagen {\O}, Denmark\\
e-mail: horsboll@alf.nbi.dk}\\
{\em and\\ $(2)$ Department of Theoretical Physics, Oxford University\\
1 Keble Road, Oxford, OX1 3NP, United Kingdom\\
e-mail: vernizzi@thphys.ox.ac.uk}\\[.5cm]

\end{center} 

\vs{10}

\centerline{ {\bf Abstract}}

\noindent
Using Random Matrix Theory we set out to compute the microscopic 
correlators of the Euclidean Dirac operator in four dimensions. 
In particular we consider: 
the chiral Or\-tho\-go\-nal Ensemble ($\chi$OE), corresponding
to a Yang-Mills theory with two colors and fermions in the fundamental
 representation, and the  chiral Symplectic Ensemble ($\chi$SE), corresponding
 to any number of colors and fermions in the adjoint representation.
In both cases we deal with an arbitrary number of massive fermions.
We use a recent method proposed by H.~Widom for deriving closed
formulas for the scalar kernels from which all spectral correlation
  functions of the $\chi$GOE and $\chi$GSE can be determined. 
Moreover, we obtain complete analytic expressions 
of such correlators in the double microscopic limit, extending previously
known results of four-dimensional QCD at $\beta=1$ and $\beta=4$ 
to the general case with $N_f$ flavors, with arbitrary quark masses and arbitrary topological charge. 
\end{titlepage}

\renewcommand{\thefootnote}{\arabic{footnote}}
\setcounter{footnote}{0}

\sect{Introduction}\label{intro}

One of the most successful and well-established physical applications of 
Random Matrix Theory (RMT) is the analysis of Quantum Chromodynamics
 (QCD) at low energies. 
In particular, the spectral statistical properties   
of the Euclidean Dirac operator in the infrared regime, can effectively be 
described within a RMT approach. 
In fact, there is a close correspondence between the finite-volume  
partition function of four dimensional QCD in the low-energy limit and 
 the partition function of a RMT with the same global symmetries 
\cite{shu93} \cite{hal95} \cite{smi95}. 
This theory is called {\em chiral} RMT ($\chi$RMT), because of its chiral 
content. 
The four-dimensional QCD Dirac operator with $N_f$ fundamental fermions 
and gauge group $SU(N_c=2)$ or $SU(N_c > 2)$, 
is described by the chiral Orthogonal ($\chi$OE) or chiral Unitary ensemble 
($\chi$UE) , respectively. 
The case of adjoint fermions 
and $SU(N_c \geq2 )$ gauge group, corresponds to 
the chiral Symplectic ensemble ($\chi$SE) \cite{ver94}. These agreements are 
valid only in the microscopic limit of $\chi$RMT,  
in which the universality-classes of the three chiral random matrix ensembles,
 manifest explicitly \cite{ver97}. 
In this particular limit, one ``magnifies'' the Dirac operator
 spectra around the origin (zero virtuality) which, 
through the Banks-Casher relation \cite{ban80}, 
reflects the existence of a non-vanishing expectation value of the chiral 
condensate 
$\Sigma = \langle \bar{\psi}\psi \rangle$, i.e.
a chiral symmetry  breaking at low energies. 
Indeed, Leutwyler and Smilga  showed \cite{leu92} that when  
$\Sigma \neq 0$ suitable spectral sum-rules must hold. Such sum-rules can be 
written as integrals over the {\em microscopic} spectral density of 
the Dirac operator \cite{shu93}, given by
\beq
\rho_{s}(z)=\lim_{V_4 \to \infty}\frac{1}{V_4\Sigma}\rho(\frac{z}{V_4\Sigma}) 
\ ,
\eeq
where $V_4$ is the Euclidean space-time volume.
The universality  of $\chi$RMT in the microscopic limit (and therefore
 the universality of  $\rho_{s}(z)$)  supports the idea
that the QCD Dirac spectra are universal in the 
large-volume scaling limit \cite{shu93}  \cite{ver97} \cite{verzah93}. 
Several facts confirms this scenario, such as analytical results and 
extensive universality studies of  microscopic spectral correlators 
\cite{ake97} \cite{dam98} \cite{gen_univ} \cite{nis98} \cite{sen98v} 
\cite{kle00} as well as
good agreement with numerical results obtained in Lattice QCD
 simulations for massless  \cite{lattice_massless} and massive 
\cite{lattice_massive} fermions.
 For more details and a complete list of references 
we suggest some excellent reviews \cite{review} \cite{verzah94}, 
in which it is also possible to find  
 information and references about non-chiral RMT and the corresponding
 three-dimensional QCD theory. \\
In the fundamental paper \cite{ake97} the large-$N$ behaviour of the 
orthogonal 
polynomials relevant for the chiral Unitary ensemble is proven to be universal 
near the origin, in the scaling limit $x=N^{2}\lambda$, where $x$ is kept fixed. From this it follows that  
all microscopic correlators have the same universal behaviour. Also 
in the so-called double microscopic limit of $\chi$UE 
where both masses and eigenvalues 
are scaled at large-$N$, one sees a similar u\-ni\-ver\-sal 
behaviour of the orthogonal polynomials associated with $N_{f}$ massive 
flavors. Consequently all 
microscopic correlators are seen to be universal \cite{dam98}. 
In the $\chi$OE and $\chi$SE 
the question whether the microscopic correlators in general are 
universal or not is still open. 
By relating the massless 
kernels of $\chi$OE and $\chi$SE to the $\chi$RMT with complex elements 
($\beta=2$), which is known to be universal, it is shown 
in ref.~\cite{sen98v}\cite{kle00} that microscopic universality 
persists under certain smoothness assumptions. Though a similar universal 
behaviour in the double microscopic limit for $\beta=1$ and 
$\beta=4$ remains to be proven, it is widely recognized that universality in 
these 
two models is a reliable consequence as well.\\
So far several results have been explicitly obtained for
 the case of massless flavors. For instance, 
the microscopic spectral density is known for all chiral ensembles in the case 
of an arbitrary number of massless flavors \cite{ver94} \cite{verzah93} 
\cite{nawa91} \cite{nafor95}. Also, 
the cases concerning an arbitrary number of massless flavors 
and an even number 
of massive flavors in non-$\chi$UE have been derived 
 in \cite{verzah94} and \cite{damnis98} respectively, 
and recently the cases of non-$\chi$OE and  non-$\chi$SE have 
been solved in \cite{hil_nic}.\\
In the case of massive flavors, the situation is as following. 
A general solution for the spectral correlators of the $\chi$UE 
is given in \cite{dam98} \cite{gu_we_wi98}. However, computing 
the microscopic spectral density in the massive (non-)$\chi$OE and massive (non-)$\chi$SE poses many difficulties arising from the use of skew-orthogonal 
polynomials and their behaviour at large-$N$. In \cite{nanis00} \cite{akkan00} 
these ensembles have been studied and a 
solution is presented for doubly and $\beta$-fold degenerate masses. Also, 
the general result of \cite{nis98} could in principle 
give the spectral density 
for all the three ensembles with an arbitrary number of flavors and
  masses, but it seems technically difficult.\\
Recently, H.~Widom introduced a new technique in \cite{Widom} for dealing with 
 OE and SE using the standard orthogonal polynomials of the UE and
 thereby avoiding skew-orthogonal polynomials. This technique 
 has already been used successfully in \cite{hil_nic} for studying 
 the massless  non-$\chi$OE and  non-$\chi$SE. 
In this paper we apply the same technique to the $\chi$OE and $\chi$SE with 
an arbirtary number of massive flavors. Applying this method presents several 
 advantages. First of all this approach seems to be a promising 
alternative to the usual one 
 with skew-orthogonal polynomials, and it gives a different point of 
 view of the general problem of determining the scalar kernels of the 
 OE and SE. Secondly, since this technique deals with standard orthogonal
 polynomials only, then it is possible in principle to get the 
microscopic limit of all the correlation functions from the 
already well-established results on the universal microscopic limit 
of orthogonal polynomials in the UE. Finally, the orthogonal polynomials
 for the general massive case 
naturally lead to a unifying notation which is
   very  helpful when used with this new technique.\\ 
This paper is organized as follows. In the next Section we shortly address 
 the definition of the chiral matrix model which is relevant for 
 four-dimensional QCD. In Section $3$ we show how it is possible
 to compute correlation functions for the $\chi$SE and $\chi$OE by means
 of the technique of Widom. Such a technique requires as a basic
 ingredient, the explicit evaluation of ortho{\em normal} polynomials for the 
general
 massive case. These are explicitly determined in Section $4$ (with their 
normalization factors) in a closed form\footnote{
The same set of polynomials have already been obtained in 
 \cite{dam98}, but only in an recursive form.}.  Within this formalism,
 the degenerate massive, and the massless case are obtained as particular
 cases of our general formulas. Moreover, we derive some equalities among
 orthogonal polynomials with different number of flavors. 
In Section $5$ we 
 apply the machinery described in Section $3$ to the case of interest here,
 and finally in Section $6$ we derive an expression for the scalar kernels
 of the massive $\chi$OE and massive $\chi$SE. The microscopic limit
 of our final formulas is discussed in Section $7$.\\ 
 Let us finally remark 
 that in this paper we do not address the question of universality 
 (which is actually considered here as a working hypothesis) of our 
 results as well as explicit numerical calculations. We 
postpone both of them, and in the following we 
focus on the application of the new technique
 by Widom to the massive case of four-dimensional QCD.

\sect{Chiral ensembles} 
A chiral random matrix model with the same symmetries 
of the  QCD partition function, can be set up by replacing 
the Dirac operator ${\mathcal D}$ 
with a suitable constant 
 off-diagonal block random matrix. The off-diagonal structure 
stems from the anticommutation relation 
between ${\mathcal D}$ and $\gamma_5$.  
Thus  $\chi$RMT for QCD in four dimensions 
is defined by the partition function \cite{shu93} \cite{ver94},
\beq
\label{full_model}
{\mathcal Z}_{\nu}^{(N_{f},\beta)}(\{m_f\})  =  \int dW 
\prod_{f=1}^{N_{f}} \det 
 \left(
{\mathcal D}+m_f
\right) 
e^{-N \frac{\beta}{2}{\rm Tr} V(W^{\dagger}W)
} \ ,
\eeq
with
\beq
{\mathcal D}=
\left(
\begin{array}{cc}
 0 & i  W  \\
i W^{\dagger} & 0 
\end{array}
\right) \ ,
\eeq
where the Dyson index $\beta=1,2,4$ labels 
the  different chiral ensembles. The rectangular
matrix $W$ of size $N \times (N+|\nu|)$ is real, complex 
or quaternion real for $\beta=1,2,4$, respectively. 
The integration measure $dW$ is the Haar measure of the group under 
consideration, the integer $\nu$ is related to the topological charge 
 \cite{hal95} \cite{leu92} and $2N+|\nu|$ is the space-time volume. 
The polynomial potential $V$ in the exponential can be replaced 
by a pure quadratic term $V(W^{\dagger}W)=W^{\dagger}W$ because of the 
universality of the matrix model in the microscopic large-$N$ limit
 \cite{ake97} \cite{sen98v}. Throughout this paper we will deal with an arbitrary topological charge 
$\nu$.\\  
The matrix integral can be rewritten, up to an irrelevant overall constant 
factor, in terms of the eigenvalues $\lambda_i$ of 
the Hermitian positive-definite matrix $W^{\dagger}W$,
 that is:
\beqn
{\mathcal Z}_{\nu}^{(N_{f},\beta)}( \{ m_f \}) & = &
\int_0^{\infty} \prod_{i=1}^{N} 
\left(
d \lambda_i \prod_{f=1}^{N_{f}} (\lambda_i+m_f^2) \, 
\lambda_i^{\beta(\nu+1)/2-1}
e^{-N \frac{\beta}{2} \lambda_i}
\right)
\left|
\Delta(\{ \lambda_k \})
\right|^{\beta} \nonumber \\
& = & \int_0^{\infty} \prod_{i=1}^{N} d \lambda_i w(\lambda_i) 
\left|
\Delta(\{ \lambda_k \})
\right|^{\beta} \ ,
\label{part_func}
\eeqn
where $\Delta(\{ \lambda_k \})=\prod_{m>n}^N(\lambda_m-\lambda_n)$ is the 
Vandermonde determinant and 
\beq
\label{weight0}
w(\lambda)=\prod_{f=1}^{N_{f}} (\lambda+m_f^2) 
\lambda^{\beta(\nu+1)/2-1} e^{-N \frac{\beta}{2} \lambda}
\eeq
 is the 
weight function on ${\mathcal I}=[0,\infty[$. These two functions 
define all the spectral statistical properties of the matrix model 
in eq.~(\ref{part_func}). The corresponding properties for the model
 in eq.~(\ref{full_model}) are expressed in terms of 
the real eigenvalues $\xi_i$ of the Dirac operator ${\mathcal D}$,
by means of $\xi^2_i=\lambda_i$.

\sect{Spectral correlation functions for $\beta=1$ and $\beta=4$ }
\label{sect_corr}
The $m$-point spectral correlation function for all three chiral 
ensembles is
\beqn
\label{corr}
R_{m}^{(\beta)}(\lambda_{1},..,\lambda_{m}) & \equiv & 
\frac{N!}{(N-m)!} \langle \prod_{i=1}^{m} {\rm Tr} \, \delta ( \lambda_i - 
W^{\dagger} W ) \rangle \\
& =& \det_{1 \leq i,j \leq m} [K^{(\beta)}_N(\lambda_i,\lambda_j)] \ ,
\label{det_quat}
\eeqn
where the expectation value in the first line, is understood with 
respect to 
the partition function in eq.~(\ref{part_func}) and the second line follows from a well-known result \cite{meh91} \cite{mha91} which expresses all the 
correlation functions
in terms of the determinant of the kernel of suitable 
polynomials\footnote{For $\beta = 1$ and $\beta=4$ the determinant in eq. 
(\ref{det_quat})
is understood as the quaternion determinant of a matrix kernel (for details 
see \cite{meh91}).}.
For $\beta=2$ such polynomials are orthogonal polynomials w.r.t. the 
weight $w(x)$ in eq. (\ref{weight0}), 
whereas when $\beta=1$ and $\beta=4$ one usually chooses 
skew-orthogonal polynomials w.r.t. $w(x)$. 
Indeed, one does not need to introduce skew-orthonormality since the 
matrix kernel is largely independent of the 
particular choice of the  polynomials \cite{tr94}. 
Therefore, if one defines 
\beqn
\label{intro_phi_beta4}
\varphi_{j}(x) & = & p_{j}(x)\ \sqrt{w(x)} \ , \quad j=0,1,\ldots,2N-1, \quad
{\rm for } \  \beta=4  \ , \\
\label{intro_phi_beta1}
\varphi_{j}(x) & = & p_{j}(x)\ w(x)  \ , \quad j=0,1,\ldots,N-1, \quad
{\rm for } \ \beta=1 \ ,
\eeqn
where $p_{j}(x)$ are arbitrary polynomials of order $j$, 
then the matrix kernels can be written as~\cite{tr98}:
\beq
\label{K4}
K_{N}^{(4)}(x,y)=\left( 
\begin{array}{cc} 
S_{N}^{(4)}(x,y) & S_{N}^{(4)} \ D(x,y) \\
I S_{N}^{(4)}(x,y) & S_{N}^{(4)}(y,x) 
\end{array} 
\right)\ ,
\eeq
and
\beq
\label{K1}
K_{N}^{(1)}(x,y)=\left( 
\begin{array}{cc} 
S_{N}^{(1)}(x,y) & S_{N}^{(1)} D(x,y) \\
I S_{N}^{(1)}(x,y)-\varepsilon(x-y) & S_{N}^{(1)}(y,x) 
\end{array} 
\right)\ ,
\eeq
where $\varepsilon(x)={\rm sgn}(x)/2$, $S^{(\beta)}_N(x,y)$ equals 
 the scalar kernel of a suitable operator $\widehat{S}^{(\beta)}_N$,
and $I S^{(\beta)}_N(x,y)$ and $D S^{(\beta)}_N(x,y)$ are the kernels 
of the operators $\widehat{I} \widehat{S}^{(\beta)}_N$,
$\widehat{D} \widehat{S}^{(\beta)}_N$ with $\widehat{I}, \widehat{D}$ identified 
as the
 integration and differentiation operators, respectively.
All the matrix elements in eq. (\ref{K4}) and eq. (\ref{K1}) are determined once 
the 
scalar kernel $S_{N}^{(\beta)}(x,y)$ is known.\\ 
Indeed, the scalar kernel $S^{(\beta)}_N(x,y)$ can be 
expressed in terms of $\varphi_j$ functions only  
and it is a matter of fact that the 
choice of skew-orthogonal polynomials leads to the simplest possible 
expressions (see for instance \cite{meh91}). 
However, skew-orthogonal polynomials are  
difficult to determine and furthermore they seem 
unsuitable for straightforward 
manipulations at large-$N$  as opposed to the standard orthogonal 
polynomials\footnote{A formula useful for studying the asymptotics of skew-orthogonal
 polynomials, has been derived recently in \cite{eynard}.}. This fact 
 has led some authors to use a different approach~\cite{Widom}. 
This approach uses orthogonal polynomials only and, without introducing
skew-orthogonal polynomials, H.~Widom derived general 
formulas for the kernels $S^{(\beta)}_N(x,y)$ expressing it as 
corrections in addition to the unitary kernel $K^{(2)}_N(x,y)$, 
with $N$ replaced by $2N$ when $\beta=4$ and $w$ replaced by 
$w^2$ when $\beta=1$. Because all quantities are 
expressed in terms of orthogonal polynomials, calculations  
and large-$N$ asymptotics are straightforward in this approach.\\
In this paper we adopt the method developed by Widom in order to 
determine the functions $S^{(\beta)}_N(x,y)$ and through that we  
analyze the original physical problem. 
In \cite{hil_nic} the same technique has already been applied successfully
 to the case of three-dimensional QCD  with $N_f$ massless fermions. 
In the remaining part of this section we briefly sketch this technique. 
For more details we refer the reader to the 
original works~\cite{Widom}~\cite{tr94} and \nopagebreak to~\cite{hil_nic}.\\

The first step of this technique, is to build-up a Hilbert
space  $\mathcal{H}$ from the functions $\varphi_j$ defined as:
\beq
\label{General_phi}
\varphi_{j}(x)  =  p_{j}(x)\ \sqrt{w(x)} \ , \quad j=0,1,\ldots,N-1 \ . \\
\eeq
According to  eq.~(\ref{intro_phi_beta4}) or eq.~(\ref{intro_phi_beta1}), 
in what follows we always assume  that 
$N$ should be replaced by $2N$ when $\beta=4$ and $w$ should be replaced by 
$w^2$ when $\beta=1$ (if not explicitly stated otherwise).  
Under suitable hypotheses on the measure\footnote{In particular
 $w'(x)/w(x)$ must be a rational function over 
$\mathcal{I}$.} $w(x)=e^{-V(x)}$, $\mathcal{H}$ is 
defined as the linear space spanned  
by the functions $\varphi_{0},...,\varphi_{N-1}$. In this picture
the scalar kernel $K_{N}^{(\beta)}(x,y)$ can be considered as the kernel 
of the projection operator\footnote{For that it is crucial that the polynomials 
$\{p_{j}(x)\}$ are  ortho{\em normal}.}
 $\widehat{K}$ onto the Hilbert space  ${\mathcal H}$, 
 and it may be written as 
\beq
\label{K}
K^{(2)}_N(x,y) = 
\sum_{j=0}^{N-1} \varphi_j(x) \varphi_j(y)= \frac{a_N}{x-y} 
\left(
\varphi_N , \varphi_{N-1}
\right)_x
\left(
\begin{array}{cc}
0 & 1 \\
1 & 0 \\
\end{array}
\right)
\left(
\begin{array}{c}
\varphi_N \\
\varphi_{N-1}\\
\end{array}
\right)_y
\ ,
\eeq
where $a_N \equiv k_{N-1}/k_N$, with $k_N$ identified as the 
highest coefficient in $p_N(x)$. 
The kernel $S_{N}^{(\beta)}(x,y)$ can be written as the unitary scalar kernel 
$K_{N}^{(2)}(x,y)$ plus extra terms, whose number is
 independent of $N$ and closely related to the commutator $[\widehat{D},
\widehat{K}]$. 
The kernel of the commutator $[\widehat{D},\widehat{K}]$ 
 is~\cite{Widom}:
\beqn
\label{DK}
[D,K](x,y) & = & (\frac{\partial}{\partial x}+\frac{\partial}{\partial 
y})K^{(2)}_{N}(x,y)\nonumber \\
& = &
a_N
\left(
\varphi_N , \varphi_{N-1}
\right)_x
\left(
\begin{array}{cc}
\frac{C(x)-C(y)}{x-y} &  \frac{A(x)-A(y)}{x-y}\\
\frac{A(x)-A(y)}{x-y} & \frac{B(x)-B(y)}{x-y} \\
\end{array}
\right)
\left(
\begin{array}{c}
\varphi_N \\
\varphi_{N-1}\\
\end{array}
\right)_y\ ,
\eeqn 
where 
\beq
A(x)=-A_N(x)-\frac{1}{2}\widetilde{V}'(x) , \quad B(x)=B_{N}(x) ,\ \textrm{and} 
\quad C(x)=\frac{a_N}{a_{N-1}}B_{N-1}(x) \ ,
\eeq
are rational functions with the same poles as the ratio $w'(x)/w(x)$, and
\beqn
\label{A_and_B_1}
A_N(x) & \equiv & a_N \int_0^{+\infty} \varphi_N(z) \, \varphi_{N-1}(z) \,
 U(x,z) \, dz \ ,\\
B_N(x) & \equiv & a_N \int_0^{+\infty} \varphi_N(z)^2 \, U(x,z) \, dz \ ,
\label{A_and_B_2}
\eeqn
with 
\beq
\label{def_u}
U(x,z)\equiv \frac{\widetilde{V}'(x)-\widetilde{V}'(z)}{x-z} \ , \quad 
\widetilde{V}'(x)=\chi(\beta)V'(x)\ ,
\eeq
and 
\beq
\label{chibeta}
\chi(\beta)= \left\{ \begin{array}{ll}
1 & \textrm{for}\ \beta=4\nonumber\\
2 & \textrm{for}\ \beta=1\nonumber 
\end{array} \right. \ ,
\eeq
stemming from the change in notation. Let 
$n_{\infty}$ and $n_{x_{i}}$ denote the pole orders of 
$w'(x)/w(x)$ at infinity and $x_{i}$, respectively. Then the functions
\beqn
\label{H1}
x^{k}\varphi_{N-1}\quad & ; & \quad x^{k}\varphi_{N} \quad (0\leq k < 
n_{\infty}) \ \ ,\\
\label{H2}
(x-x_{i})^{-k-1}\varphi_{N-1}\quad & ; & \quad (x-x_{i})^{-k-1}\varphi_{N} \quad 
(0\leq k < n_{x_{i}}) \ \ ,
\eeqn
span a subspace $\mathcal{H}_{\mathrm{sub}} \subset \mathcal{H}$ 
and a corresponding subspace $\mathcal{H}_{\mathrm{sub}}^{\perp} \subset 
\mathcal{H}^{\perp}$
both of dimension $n=n_{\infty}+\sum_{i}n_{x_{i}}$.
 The subspace $\mathcal{H}_{\mathrm{sub}}$ 
is determined by the condition  that 
$\mathcal{H}_{\mathrm{sub}} \subset \mathcal{H}$, whereas 
the subspace $\mathcal{H}_{\mathrm{sub}}^{\perp}$
is determined by the $n$ orthogonality conditions set up by the functions 
\beq
\varphi_{N-k}\quad (k<n_{\infty})\quad ; \quad\varphi_{k}\quad 
(k<n-n_{\infty})\quad .
\eeq
Let $\psi_{1},...,\psi_{n}\in \mathcal{H}$ and $\psi_{n+1},...,\psi_{2n}\in 
\mathcal{H}^{\perp}$ denote these $2n$ linearly independent functions, then one 
has
\beq
\label{Aij}
[D,K](x,y)=\sum_{i,j=1}^{2n}\psi_{i}(x)A_{ij}\psi_{j}(y)\ ,
\eeq
and once the $\psi$'s are chosen, the symmetric constant matrix $A=[A_{ij}]$ is 
uniquely determined. Furthermore, the matrix $A$ is always in a  
block off-diagonal form:
\beq
A_{ij}=0 \quad i,j \leq n \quad \textrm{or}\quad i,j>n\ .
\eeq
One also defines the matrices:
\beqn
\label{Bij}
B_{ij}=(\hat{\varepsilon}\psi_{i},\psi_{j}) & = & 
\int_{\mathcal I} \int_{\mathcal I} dx dy \,  \varepsilon(x-y) \,
 \psi_i(x) \,  \psi_j(y) \ , 
\\
J & = &
\left(
\begin{array}{ccc}
I_{n \times n} & \vdots & 0_{n \times n} \\
\cdots & \cdots & \cdots \\
0_{n \times n} & \vdots & 0_{n \times n} 
\end{array}
\right)_{2n \times 2n} \ , \nonumber \\
C & = & J+B A \ , \label{Cmatrix} 
\eeqn
with $0_{n \times n}$, $I_{n \times n}$ being the $n \times n$ null matrix 
and the identity matrix, respectively. The operator $\hat{\varepsilon}$ 
is defined as
\beq
\label{epspsi}
(\hat{\varepsilon}\psi_{i})(x)= \int_{\mathcal I} dx  \varepsilon(x-y) \psi_i(x) 
\ .
\eeq
Defining $A_{0}, C_{0}\ \textrm{and}\ C_{00}$ as the matrices 
obtained by deleting from the corresponding matrices 
the last $n$ columns, the last $n$ rows and the last $n$ rows and 
 columns respectively, one finally has the main result of~\cite{Widom}, 
that is 
\beqn
\label{SS1}
S_{N}^{(1)}(x,y) & = & K^{(2)}_{N}(x,y)-\sum_{i\leq n,j=1}^{2n} 
[AC(I-BAC)^{-1}]_{ji}
\psi_{i}(x)\varepsilon\psi_{j}(y) \ , \\
\label{SS4}
S_{N}^{(4)}(x,y) & = & 
K^{(2)}_{N}(x,y)-\sum_{i>n,j=1}^{2 
n}[A_{0}C_{00}^{-1}C_{0}]_{ij}\psi_{i}(x)\varepsilon\psi_{j}(y)\ ,
\eeqn
remembering once again that $N \to 2N$ in eq. (\ref{SS4}) 
and $w\to w^{2}$ in eq. (\ref{SS1}).\\ 
A remarkable feature of eq. (\ref{SS1})
 and eq. (\ref{SS4}) is that the scalar kernels $S_{N}^{(\beta)}(x,y)$ 
naturally are expressed as a {\em finite} number of corrections
 to the unitary kernel $K_{N}^{(2)}(x,y)$, a number which is independent
 of $N$. This fact will prove its importance in Section \ref{sectML} where
 we study the microscopic limit of the quantities of interest here. 
Finally, inserting eq. (\ref{SS1}) and eq. (\ref{SS4}) into 
eq. (\ref{K1}) and eq. (\ref{K4}), one can calculate eq. (\ref{det_quat}) 
and through that all possible spectral correlation functions defined by 
eq. (\ref{corr}).\\

\sect{Orthogonal polynomials for $N_f$ massive fermions}

In this section we apply the technique of Widom, described above,   
to the general massive chiral case with $N_f$ flavors.
The first ingredient we are looking for are polynomials 
$P_n^{(N_f,\alpha)}(x;m_1,\ldots,m_{N_f})$ ortho{\it normal}  
with respect to the weight function
\beq
\label{weight}
w^{(N_f,\alpha)}(x) \equiv \prod_{i=1}^{N_f} (x+m_i^2) \, x^{\alpha} \, e^{-x} 
\eeq
defined on the interval ${\cal I}=[0,+\infty[$, where $\alpha$ is a real 
non-negative
constant. In this formalism, it is understood that 
$w^{(0,\alpha)}(x) \equiv x^{\alpha} e^{-x} $. The weight in 
eq. (\ref{weight}) is seen to be slightly different from the one in eq. 
(\ref{weight0}),
 but if we identify $\alpha=\beta (\nu+1)/2-1$ and $c=N \beta/2$, 
then we observe that 
\begin{itemize}
\item the variable $\alpha$ could be negative (e.g. $\alpha=-1/2$ 
for  $\beta=1$ and $\nu=0$). Although we suppose $\alpha \geq 0$ 
throughout this paper, the final results are valid also for $\alpha>-1$ 
 {\em via} analytic continuation \cite{Widom};\
\item the exponential function in eq. (\ref{weight}) should be $e^{-cx}$ indeed. 
 Here we suppose $c=1$ instead, but at the end of Section 
\ref{kernelS} we shall 
 extend all the results to the general case $c \neq 1$, by means of scaling 
arguments.
\end{itemize}
In order to determine the polynomials $P_n^{(N_f,\alpha)}$ explicitly, 
we fix some useful
notation at first. We define the scalar product 
 of two real functions $f(x)$, $g(x)$ with respect 
to the weight function $w^{(N_f,\alpha)}(x)$ by  
\beq
\langle f,g \rangle_{N_f,\alpha} \equiv \int_{\cal I} dx \, 
w^{(N_f,\alpha)}(x) f(x) \, g(x) \ ,
\eeq
and furthermore we shall use also the following shortened and suggestive  
notation:
\beqn
P_n^{(N_f,\alpha)}(x;{\bf m}) & \equiv &  
P_n^{(N_f,\alpha)}(x;m_1,\ldots,m_{N_f}) \nonumber \ , \\
P_n^{(N_f,\alpha)}(x;{\bf m},0) & \equiv &  
P_n^{(N_f,\alpha)}(x;m_1,\ldots,m_{N_f-1},0) \nonumber \ ,  \\
P_n^{(N_f-1,\alpha)}(x;{\bf m}_{\neq i}) & \equiv &  
P_n^{(N_f-1,\alpha)}(x;m_1,\ldots,m_{i-1},m_{i+1},\ldots,m_{N_f}) \nonumber 
\ . \eeqn
Moreover, let $k^{(N_f,\alpha)}_n$ 
denote the highest coefficient 
in $P_n^{(N_f,\alpha)}=k^{(N_f,\alpha)}_n x^n+\ldots$,
  and let us define the $(N_f+1) \times (N_f+1)$ matrix 
\beq
\label{Lambda}
\Lambda^{(N_f,\alpha)}_n(x) \equiv \left(
\begin{array}{llcl}
P^{(0,\alpha)}_n(x) & P^{(0,\alpha)}_{n+1}(x) & \cdots & 
P^{(0,\alpha)}_{n+N_f}(x) \\
P^{(0,\alpha)}_n(-m_1^2) & P^{(0,\alpha)}_{n+1}(-m_1^2) & 
\cdots & P^{(0,\alpha)}_{n+N_f}(-m_1^2) \\
\cdots & \cdots & \cdots & \cdots \\
P^{(0,\alpha)}_n(-m_{N_f}^2) & P^{(0,\alpha)}_{n+1}(-m_{N_f}^2)& \cdots 
& P^{(0,\alpha)}_{n+N_f}(-m_{N_f}^2) \\
\end{array}
\right) \ .
\eeq
Finally, we identify $\Lambda^{(N_f,\alpha)}_{n,i}$, $i=1,\ldots,N_f+1$, 
as the $N_f \times N_f$ sub-matrix   
obtained by omitting the first row and the
$i$-th column from $\Lambda^{(N_f,\alpha)}_n(x)$ 
in the definition (\ref{Lambda}). Notice that all $\Lambda$ matrices are expressed
 in terms of orthonormal polynomials $P_i^{(0,\alpha)}(x)$
 with $N_f=0$ flavors only. Since the polynomials $P_n^{(0,\alpha)}(x)$ are 
orthonormal w.r.t. the measure $w^{(0,\alpha)}(x)$, then they are necessarily 
proportional to the generalized Laguerre 
polynomials $L^{(\alpha)}_n(x)$ \cite{Abramo}, i.e. 
\beq
\label{pn0}
P_n^{(0,\alpha)}(x)=L^{(\alpha)}_n(x) / \sqrt{h^{\alpha}_n} \ \ ,
 \quad   h^{\alpha}_n=\frac{\Gamma(n+\alpha+1)}{n!} \, .
\eeq
Therefore, from $L^{(\alpha)}_n(x)=(-1)^n x^n/n!+\ldots$ one reads
the highest coefficients
\beq
\label{high_lagrre}
k^{(0,\alpha)}_n=\frac{(-1)^n}{\sqrt{\Gamma(n+\alpha+1) \, n!}} \ .
\eeq
Now, it is well-known that the orthonormality condition
\beq
\label{orthonorm}
\langle
P_i^{(N_f,\alpha)}(x;{\bf m}) , P_j^{(N_f,\alpha)}(x;{\bf m})
\rangle_{N_f,\alpha}
=\delta_{ij}
\eeq
uniquely determines polynomials $P_i^{(N_f,\alpha)}$, up to a relative 
sign. Under the hypothesis that all the masses $\{ m_i \}$ are distinct, these 
polynomials can, according to Christoffel's theorem \cite{Szego}, be represented 
in terms of $P_n^{(0,\alpha)}(x)$ as follows: 
\beq
\label{CT}
P_n^{(N_f,\alpha)}(x;{\bf m})
= \frac{1}{\sqrt{h^{(N_f,\alpha)}_n({\bf m})}}
\frac{\det \left[ \Lambda^{(N_f,\alpha)}_{n}(x) \right]}
{\prod_{i=1}^{N_f} (x+m_i^2)} \ .
\eeq
The normalization factor $h^{(N_f,\alpha)}_n$ is 
\beq
\label{norm}
h^{(N_f,\alpha)}_n({\bf m})=(-1)^{N_f} 
\frac{k^{(0,\alpha)}_{n+N_f}}{k^{(0,\alpha)}_n} \det \left[ 
\Lambda^{(N_f,\alpha)}_{n,1} \Lambda^{(N_f,\alpha)}_{n,N_f+1} \right] 
\eeq
and the highest coefficient is 
\beq
\label{knnf}
k^{(N_f,\alpha)}_n({\bf m}) =  \sqrt{ (-1)^{N_f} 
 k_n^{(0,\alpha)}  k_{n+N_f}^{(0,\alpha)} 
\frac{\det \left[ \Lambda^{(N_f,\alpha)}_{n,N_f+1}  \right]}
{\det \left[ \Lambda^{(N_f,\alpha)}_{n,1}  \right]}  } 
\ .
\eeq
where $k_n^{(0,\alpha)}$ is the coefficient given in eq. (\ref{high_lagrre}). 
For an explicit derivation of eq. (\ref{norm}) and (\ref{knnf}) we refer to 
appendix \ref{app_norm}. 
Eq. (\ref{CT}) is completely symmetric under permutations of the 
masses: although the determinant in the numerator is completely
antisymmetric, the algebraic square root in the denominator is also completely 
antisymmetric (with $\sqrt{(-1)^2}=-1$). This observation is consistent 
with the fact that also the weight function in eq. (\ref{weight}) is completely
 symmetric under permutations of the masses. Moreover,  
the case of a degenerate mass $m_k$ is understood in a limit sense 
of eq. (\ref{CT}). For instance, 
if there are only two degenerate masses $m_1=m_2$, we just take the limit 
of eq. (\ref{CT}) as $m_2 \to m_1$. Dividing both numerator and denominator
by $m_2^2-m_1^2$, and substituting 
the third row $\{ P_i^{(0,\alpha)}(-m_2^2) \} $ 
with the combination  
 $ \{ P_i^{(0,\alpha)}(-m_2^2)-P_i^{(0,\alpha)}(-m_1^2) \}$, 
 in all the determinants, we obtain in the limit $m_1 \to m_2$, that 
the third row effectively is replaced 
by the first-order derivatives $d P_i^{(0,\alpha)}(y)/dy$ at 
$y=-m^2_1$. In the case of a degenerate mass $m_k$ of multiplicity 
$l$, $l>1$, we replace in all the determinants of eq. (\ref{CT})  
the corresponding rows $k+1,k+2,\ldots, k+l$ with the 
derivatives of order $0,1,2,\ldots,l-1$ of the 
polynomials $P_i^{(0,\alpha)}(y)$ evaluated at $y=-m_k^2$. It is
 worthwhile to remind here that degenerate masses are definitely  
 needed for the $\beta=1$ case, because $w \to w^2$ 
effectively implies $N_f \to 2 N_f$, i.e. $ \{ m_1,m_2, \ldots ,m_{N_f} \} \to
 \{ m_1,m_1,m_2,m_2, \ldots ,m_{N_f},m_{N_f}  \} $.\\

Let us give two explicit examples where eq. (\ref{CT}) is applied. 
In the case $N_f=1$, it reads
\beq
\label{poly_nf1}
P_n^{(1,\alpha)}(x;m_1) = \sqrt{ \frac{n! (n+1)! }
{\Gamma(n+\alpha+1) \Gamma(n+\alpha+2)}}
\frac{ 
L_n^{(\alpha)}(x) L_{n+1}^{(\alpha)}(-m_1^2)-L_{n+1}^{(\alpha)}(x) 
L_n^{(\alpha)}(-m_1^2) }
{(x+m_1^2) \sqrt{h^{(1,\alpha)}_n(m_1)}}
\eeq
where 
\beq
h^{(1,\alpha)}_n(m_1)=\frac{n!}{\Gamma(n+\alpha+2)} \left[ 
L_{n+1}^{(\alpha)}(-m_1^2)  L_n^{(\alpha)}(-m_1^2) \right] \ .
\eeq
In the case $N_f=2$, with degenerate masses $m_1=m_2=m$, it reads
\beq
\label{poly_nf2}
P_n^{(2,\alpha)}(x;m,m)=\frac{\det
\left(
\begin{array}{lll}
L_n^{(\alpha)}(x) & L_{n+1}^{(\alpha)}(x) & L_{n+2}^{(\alpha)}(x) \\
L_{n}^{(\alpha)}(-m^2)  & L_{n+1}^{(\alpha)}(-m^2) & L_{n+2}^{(\alpha)}(-m^2) \\
L_n^{(\alpha)'}(-m^2) &  L_{n+1}^{(\alpha)'}(-m^2) & L_{n+2}^{(\alpha)'}(-m^2) \\
\end{array}
\right)
}{(x+m^2)^2 \sqrt{h_n^{\alpha}  
h_{n+1}^{\alpha} h_{n+2}^{\alpha} h_n^{(2,\alpha)}(m,m)} }
\eeq
and 
\beq
h_n^{(2,\alpha)}(m,m)=
\frac{n! (n+1)! 
\det
\left(
\begin{array}{ll}
 L_{n+1}^{(\alpha)}(-m^2) & L_{n+2}^{(\alpha)}(-m^2) \\
 L_{n+1}^{(\alpha)'}(-m^2) & L_{n+2}^{(\alpha)'}(-m^2) 
\end{array}
\right)
\left(
\begin{array}{ll}
L_{n}^{(\alpha)}(-m^2)  & L_{n+1}^{(\alpha)}(-m^2) \\
L_{n}^{(\alpha)'}(-m^2) &  L_{n+1}^{(\alpha)'}(-m^2) \\
\end{array}
\right)
}
{\Gamma(n+\alpha+2) \Gamma(n+\alpha+3)} \ . 
\eeq
All the derivatives can easily be evaluated using the property 
$dL_n^{(\alpha)}(x)/dx=-L_{n-1}^{\alpha+1}$, $n>0$ 
(iteratively, if needed).\\
These examples are just two particular cases of a general fact: 
 all the orthogonal polynomials for the general massive 
case are 
expressed nicely in terms of generalized Laguerre polynomials by means of 
eq. (\ref{CT}), (\ref{norm}) and (\ref{high_lagrre}).\\  
Now, for the sake of future use we point out two remarkable properties 
of the polynomials $P_n^{(N_f,\alpha)}$. 
\begin{enumerate}
\item Let us consider the set $\{m_1,\ldots, 
m_{N_f},0 \}$ consisting of $N_f+1$ flavors. In this case eq. (\ref{orthonorm}) 
reads
\beq
\delta_{ij}=
\left.
\langle
P_i^{(N_f+1,\alpha)}(x;{\bf m},0) ,
P_j^{(N_f+1,\alpha)}(x;{\bf m},0)
\rangle_{N_f+1,\alpha} \right|_{m_{N_f+1}=0}
\eeq
that is, the polynomials 
$P_n^{(N_f+1,\alpha)}(x;{\bf m},0)$
are orthonormal with respect to the measure 
$x^{\alpha+1} \prod_{i=1}^{N_f}(x+m_i^2) e^{-x} dx$. On 
the other hand, also the polynomials 
$P_n^{(N_f,\alpha+1)}(x;{\bf m})$ 
are orthonormal w.r.t the same measure, i.e.
\beq
\delta_{ij}=
\langle
P_i^{(N_f,\alpha+1)}(x;{\bf m}),
P_j^{(N_f,\alpha+1)}(x;{\bf m})
\rangle_{N_f,\alpha+1} \ .
\eeq
From uniqueness of orthogonal polynomials w.r.t. a given measure
it follows
\beq
\label{diag_move}
P_n^{(N_f+1,\alpha)}(x;{\bf m},0)=
P_n^{(N_f,\alpha+1)}(x;{\bf m}) \ ,
\eeq
for every $n$ and $N_f$.\footnote{A trivial check of formula 
(\ref{diag_move}) is for $N_f=0$. Indeed, if we put $m_1=0$ in 
eq. (\ref{poly_nf1}) then we obtain exactly $P_n^{(0,\alpha+1)}(x)$, 
eq. (\ref{pn0}). Similarly, if we put $m=0$ in eq. (\ref{poly_nf2})
 then we obtain 
$P_n^{(2,\alpha)}(x;0,0)=P_n^{(1,\alpha+1)}(x;0)=P_n^{(0,\alpha+2)}(x)$.}
\\ 
\item  Christoffel's theorem eq. (\ref{CT}) can also be  
 stated as
\beq
\label{CT_recurr}
P_n^{(N_f,\alpha)}(x;{\bf m}) = 
c_{n,i}^{(N_f,\alpha)}({\bf m})
\frac{ \det \left(
\begin{array}{ll}
P_n^{(N_f-1,\alpha)}(x;{\bf m}_{\neq i}) & 
P_{n+1}^{(N_f-1,\alpha)}(x;{\bf m}_{\neq i}) \\
P_n^{(N_f-1,\alpha)}(-m_i^2;{\bf m}_{\neq i}) &
P_{n+1}^{(N_f-1,\alpha)}(-m_i^2;{\bf m}_{\neq i})
\end{array}
\right) }{(x+m_i^2)}  \ ,
\eeq
for $i=1,\ldots,N_f$, where the coefficient $c_{n,i}^{(N_f,\alpha)}$  
is easily determined by comparing the highest coefficients on the two sides of eq. 
(\ref{CT_recurr}), i.e.
\beq
\label{CT_coeff}
c_{n,i}^{(N_f,\alpha)}({\bf m})=-\frac{ k_{n}^{(N_f,\alpha)}({\bf m})}
{k_{n+1}^{(N_f-1,\alpha)}({\bf m}_{\neq i})} \frac{1}
{P_n^{(N_f-1,\alpha)}(-m_i^2;{\bf m}_{\neq i})} \ .
\eeq
And, by means of the Christoffel-Darboux 
formula \cite{Szego}, 
one can equivalently write eq. (\ref{CT_recurr}) as 
\beq
\label{c-darboux}
P_n^{(N_f,\alpha)}(x;{\bf m})= d_{n,i}^{(N_f,\alpha)}({\bf m}) 
\sum_{j=0}^{n} 
P_j^{(N_f-1,\alpha)}(-m_i^2;{\bf m}_{\neq i})
P_j^{(N_f-1,\alpha)}(x;{\bf m}_{\neq i}) \ ,
\eeq
for $i=1,\ldots,N_f$ where the coefficient 
$d_{n,i}^{(N_f,\alpha)}({\bf m})$
is determined as before by comparing the highest coefficients in 
eq. (\ref{c-darboux}), i.e.
\beq
\label{dcoeff_1}
d_{n,i}^{(N_f,\alpha)}({\bf m})=\frac{k_n^{(N_f,\alpha)}({\bf m})}
{ k_{n}^{(N_f-1,\alpha)}({\bf m}_{\neq i})} \frac{1}
{P_n^{(N_f-1,\alpha)}(-m_i^2;{\bf m}_{\neq i})} \ .
\eeq
The coefficients $k^{(N_f,\alpha)}_n$'s which appear 
both in eq. (\ref{CT_coeff}) and 
in eq. (\ref{dcoeff_1}), are given explicitly in eq. (\ref{knnf}).
\end{enumerate}
These two considerations will be useful in the evaluation of some integrals 
appearing in next Section.\\

\sect{The Hilbert space ${\mathcal H}$}
\label{sect_hilbert}
Once the orthonormal polynomials are known then 
the next step is to build-up 
a suitable Hilbert space according to the technique 
described in Section \ref{sect_corr}. In order to do that, from the  
orthonormal polynomials $P_n^{(N_f,\alpha)}(x;{\bf m})$ 
we define the functions $\varphi_j(x)$ as in eq.~(\ref{General_phi}):
\beq
\label{phi}
\varphi_j(x) \equiv  P_j^{(N_f,\alpha)}(x;{\bf m}) \sqrt{w^{(N_f,\alpha)}(x)} 
\quad , \ \ j=0,1,\ldots \quad .
\eeq
The Hilbert space is ${\cal H}={\rm span} \{ \varphi_0,\varphi_1,\ldots,
\varphi_{N-1} \} $, and the kernel of the projection operator $\widehat{K}$ 
onto ${\mathcal H}$  and the kernel 
of the operator $[\widehat{D},\widehat{K}]$ are given in eq. (\ref{K}) 
 and eq. (\ref{DK}), respectively. In particular, we notice that the former
 can be nicely written in a very compact form:
\beqn
\label{Kcompact}
K^{(2)}_N(x,y) & = &\sum_{j=0}^{N-1} 
 P_j^{(N_f,\alpha)}(x;{\bf m}) 
 P_j^{(N_f,\alpha)}(y;{\bf m}) 
\sqrt{w^{(N_f,\alpha)}(x) w^{(N_f,\alpha)}(y)} \\
& = & \frac{ P_{N-1}^{(N_f+1,\alpha)}(x;{\bf m},\sqrt{-y})}{d_{N-1,N_f+1}^{
(N_f+1,\alpha)}({\bf m}, \sqrt{-y}) }
\sqrt{w^{(N_f,\alpha)}(x) w^{(N_f,\alpha)}(y)} \ ,
\eeqn
where we used eq. (\ref{c-darboux}) in the second equality. \\
Since $w^{(N_f,\alpha)}(x) \equiv e^{-V(x)}$, i.e.
\beq
V(x)= x-\alpha \log x -\sum_{i=1}^{N_f} 
\log(x+m_i^2) \ ,
\eeq
eq. (\ref{def_u}) reads
\beq
\label{U_xz}
U(x,z)=\chi (\beta) \left[ \frac{\alpha}{xz} +\sum_{i=1}^{N_f} 
\frac{1}{(x+m_i^2)(z+m_i^2)} \right] \ .
\eeq
where $\chi (\beta) $ is defined in eq. (\ref{chibeta}). 
Therefore, substituting eq. (\ref{U_xz}) and (\ref{phi}) into 
eq. (\ref{A_and_B_1}) and eq. (\ref{A_and_B_2}) 
 we obtain the matrix elements in eq. (\ref{DK}):
\beqn
\frac{A(x)-A(y)}{x-y} & = & a_N \chi(\beta) \left[ 
\frac{\alpha}{xy}  
\left( I^{(N,N-1)}_0-\frac{1}{2 a_N} \right)
+\sum_{i=1}^{N_f}\frac{I^{(N,N-1)}_i-\frac{1}{2 a_N}}
{(x+m_i^2)(y+m_i^2)} \right] \ , \label{AA} \\
\frac{B(x)-B(y)}{x-y} & = & \frac{B_N(x)-B_N(y)}{x-y} = -a_N  \chi(\beta)
\left[ 
\frac{\alpha I^{(N,N)}_0}{xy}+
\sum_{i=1}^{N_f}\frac{I^{(N,N)}_i}{(x+m_i^2)(y+m_i^2)} 
\right]   \ ,  		\label{BB}   \\
\frac{C(x)-C(y)}{x-y} & = & \frac{a_N}{a_{N-1}} 
\frac{B_{N-1}(x)-B_{N-1}(y)}{x-y} \ ,  \label{CC}
\eeqn
where $a_N=k_{N-1}^{(N_f,\alpha)}({\bf m})/k_{N}^{(N_f,\alpha)}({\bf m})$
 and the constants $I^{(N,j)}_0$, $I^{(N,j)}_i$  are defined as
\beqn
I^{(N,j)}_0 &=& \langle  
P_N^{(N_f,\alpha)},\frac{P_j^{(N_f,\alpha)}}{z} 
\rangle_{N_f,\alpha}  \label{IN0} \ , \\ 
I^{(N,j)}_i &=& \langle 
P_N^{(N_f,\alpha)},\frac{P_j^{(N_f,\alpha)}}{z+m_i^2} 
\rangle_{N_f,\alpha} \ , \quad i=1,2,\ldots,N_f \quad . \label{INj}
\eeqn
The above integrals are evaluated in appendix \ref{app_int}
using the previously stated properties of orthonormal polynomials
 for general $N_f$. Thus we have 
\beqn
I^{(N,N-1)}_0 & = & d_{N,N_f+1}^{(N_f+1,\alpha-1)}({\bf m},0) 
P_{N-1}^{(N_f,\alpha)}(0;{\bf m})   \nonumber \\
I^{(N,N)}_0   & = & d_{N,N_f+1}^{(N_f+1,\alpha-1)}({\bf m},0) 
P_{N}^{(N_f,\alpha)}(0;{\bf m}) \nonumber \\
I^{(N,N-1)}_i & = & d_{N,i}^{(N_f,\alpha)}({\bf m})
P_{N-1}^{(N_f,\alpha)}(-m_i^2;{\bf m}) \nonumber \\
I^{(N,N)}_i   & = & d_{N,i}^{(N_f,\alpha)}({\bf m})
P_{N}^{(N_f,\alpha)}(-m_i^2;{\bf m}) \nonumber 
\eeqn
which appear in eq. (\ref{AA}), (\ref{BB}) and (\ref{CC}).\\

\noindent
Now we can construct the $\psi_i(x)$ functions.
Since the logarithmic derivative of the weight is a rational function
\beq
\frac{dw^{(N_f,\alpha)}(x)/dx}{w^{(N_f,\alpha)}(x)}=
\frac{\alpha}{x} + \sum_{i=0}^{N_f} \frac{1}{x+m_i^2}-1 \ ,
\eeq
with no poles at infinity and  
simple poles at $x=0$ (for $\alpha \neq 0$) and at $x=-m_i^2$, 
$i=1,\ldots,N_f$, we have that the subspace 
${\mathcal H}_{sub}$ has dimension $n=N_f+1$. The number of 
linearly independent functions $\psi_k \in {\cal H}$ we are 
looking for is thus equal to $n=N_f+1$. According to 
eq. (\ref{H1}) and (\ref{H2}) we choose the following
linear combinations:
\beq
\label{psi}
\psi_i(x)  = \frac{\varphi_N^{(N_f,\alpha)}(x)}{x+\sigma_i^2}+
D_i\frac{\varphi_{N-1}^{(N_f,\alpha)}(x)}{x+\sigma_i^2} \ , \quad
i=1,\ldots,N_f+1 \ ,
\eeq
where we used the shortened notation 
 $\sigma_i \equiv m_i$ for $i=1,\ldots, N_f$ and $\sigma_{N_f+1} \equiv 0$. 
We determine all the coefficients $D_i$ by requiring 
that the functions $\psi_i$
are non singular at $x=-\sigma_i^2$, since they
have to belong to the Hilbert space ${\cal H}$. This means 
that the numerator in eq. (\ref{psi}) should be vanishing at 
$x=-\sigma_i^2$, i.e.
\beq
D_i=-\frac{P_N^{(N_f,\alpha)}
(-\sigma_i^2;{\bf m})}{P_{N-1}^{(N_f,\alpha)}(-\sigma_i^2;{\bf m})} \ , \quad 
i=1, \ldots N_f+1 \ \ .
\label{D}
\eeq
Thus all the functions $\psi_i \in {\cal H}$ 
are determined completely. In appendix \ref{app_Dcoef} we obtain 
the following compact final expression:
\beq
\label{psii1}
\psi_i(x)=k_N^{(N_f,\alpha)}({\bf m})
\frac{P_{N-1}^{(N_f+1,\alpha)}(x;{\bf m},\sigma_i)}
{k_{N-1}^{(N_f+1,\alpha)}({\bf m},\sigma_i)} \sqrt{w^{(N_f,\alpha)}(x)}
 \ , \quad i=1,\ldots,N_f+1 \ .
\eeq 

We now look for the functions 
$\psi_{N_f+2},\ldots,\psi_{2N_f+2} \in 
{\cal H}^{\perp}$. We choose again to consider the combination
\beq
\label{Dperp}
\psi_{N_f+1+i}(x) =  \frac{\varphi_N^{(N_f,\alpha)}(x)}{x+\sigma_i^2}+
D^{\perp}_i\frac{\varphi_{N-1}^{(N_f,\alpha)}(x)}{x+\sigma_i^2} 
\ ,  \quad i=1,\ldots,N_f+1 \ ,
\eeq
where $D^{\perp}_i$ are coefficients to be determined.
As discussed in Section \ref{sect_corr} the space 
${\cal S}={\rm span} \{ \varphi_0,\ldots,\varphi_{N_f} \}$ 
is a  $n$-dimensional subspace of ${\cal H}$, 
hence\footnote{This fact implies $N \geq n=N_f+1$, 
because ${\mathcal H}$ is $N$-dimensional.} we can fix all the 
$D^{\perp}_i$'s 
by imposing that each $\psi_{N_f+1+i} \in {\cal H}^{\perp}$ is orthogonal 
to all the $ \{ \varphi_0,\ldots,\varphi_{N_f} \} $, i.e.
\beq
\label{costraint}
\int_{0}^{+\infty} \psi_{N_f+1+i}(x) \, \varphi_j(x) =0 \ ,
\eeq
for $i=1,\ldots,N_f+1$ and $j=0,\ldots, N_f$. When $i=1,\ldots,N_f$,
 by substituting eq. (\ref{Dperp}) and (\ref{phi}) in the 
constraint equation (\ref{costraint}), we see that all the integrals 
are of the form eq. (\ref{IN0}) or (\ref{INj}), 
which have already been solved explicitly in eq.~(\ref{I1}) and (\ref{I0}).  
Therefore, the constraint conditions (\ref{costraint}) can be written as:
\beq
\label{dperpcoeff}
D_i^{\perp}=-\frac{d_{N,i}^{(N_f,\alpha)}({\bf m})}
{d_{N-1,i}^{(N_f,\alpha)}({\bf m})}  \ , \quad i=1,\ldots,N_f \, ,
\ ,
\eeq
with $N \geq N_f+1$. 
Notice that the bound  $N \geq N_f+1$ is fully consistent 
with our purpose of taking the microscopic large-$N$ 
limit (see Section \ref{sectML}).
When $i=N_f+1$ 
we have 
\beq
\label{dperpcoeffbis}
D_{N_f+1}^{\perp}=-
\frac{d_{N,N_f+1}^{(N_f+1,\alpha-1)}({\bf m},0)}
{d_{N-1,N_f+1}^{(N_f+1,\alpha-1)}({\bf m},0)} \ .
\eeq
In appendix \ref{app_Dcoef} we obtain a more compact and explicit 
 expression for $\psi_{N_f+1+i}$, i.e.
\beq
\label{psii21}
\psi_{N_f+1+i}(x) 
=k_{N}^{(N_f,\alpha)}({\bf m})
\frac{P_N^{(N_f-1,\alpha)}(x;{\bf m}_{\neq i})}
{k_{N}^{(N_f-1,\alpha)}({\bf m}_{\neq i})} 
\frac{\sqrt{w^{(N_f,\alpha)}(x)}}{(x+m_i^2)} \ , 
\eeq
for $i=1,\ldots,N_f$ and
\beq
\label{psii22}
\psi_{2N_f+2}(x) = 
k_{N}^{(N_f,\alpha)}({\bf m})
\frac{P_N^{(N_f,\alpha-1)}(x;{\bf m})} 
{k_{N}^{(N_f,\alpha-1)}({\bf m})}
\frac{\sqrt{w^{(N_f,\alpha)}(x)}}{x} \ ,
\eeq
for $i=N_f+1$. \\

\sect{The kernels $S_N^{(\beta)}(x,y)$}
\label{kernelS}

Now we have all the necessary ingredients 
to determine the symmetric $2n \times 2n$ 
matrix $A$, which is defined by eq. (\ref{Aij}).
In \cite{Widom} it is proven that the matrix $A$  always has the 
block form
\beq
A=\left(
\begin{array}{ccc}
0_{n \times n} & \vdots & \bar{A}_{n \times n} \\
\cdots & \cdots & \cdots \\
\bar{A}^{\dag}_{n \times n} & \vdots & 0_{n \times n} \\
\end{array}
\right)_{2n \times 2n} \ ,
\eeq 
where $0_{n \times n}$ is the $n \times n$ null matrix. 
With our choice of the $\psi$ functions, 
the $n \times n$ matrix $\bar{A}$ is always diagonal. Indeed, 
it is sufficient to prove that the off-diagonal terms 
$\bar{A}_{ij}$ are identically zero. In fact, 
 if such terms were not zero then they would be responsible for having 
mixed terms in $[D,K](x,y)$, eq. (\ref{Aij}), of the type  
$\psi_i(x) \psi_{j \neq i}(y)$, that is
mixed terms of the type $1/[x(y+m_i^2)]$ or $1/[(x+m_j^2)(y+m_i^2)]$ 
($i \neq j$). But such mixed terms are definitely not there 
in eq. (\ref{AA}), (\ref{BB}) and (\ref{CC}) and therefore, through 
eq. (\ref{DK}), they do not appear at all in $[D,K](x,y)$.
Finally, from eq. (\ref{Aij}), 
we necessarily conclude that $\bar{A}_{ij}=0$ for $i \neq j$.\\
Let us define $\bar{A}_i 
\equiv \bar{A}_{ii}$ as the diagonal elements of $\bar{A}$,
 $i=1,\ldots,N_f+1$. These elements are uniquely determined by 
comparing eq. (\ref{DK}) and eq. (\ref{Aij}): indeed, in our case it 
is sufficient to compare the highest coefficients in $x$ and $y$. 
In particular, in eq. (\ref{DK}) the terms proportional
to $P^{(N_f,\alpha)}_N(x) P_N^{(N_f,\alpha)}(y)$ are contained in
\beq
a_N P^{(N_f,\alpha)}_N(x) \frac{C(x)-C(y)}{x-y} P^{(N_f,\alpha)}_N(y) \ ,
\label{PCCP}
\eeq
whereas in eq. (\ref{Aij}) and (\ref{psi}) such terms are contained in 
\beq
2 P^{(N_f,\alpha)}_N(x) \left[ \sum_{i=1}^{N_f} 
\frac{\bar{A}_i}{(x+m_i^2)(y+m_i^2)}
+\frac{\bar{A}_{N_f+1}}{xy} \right] P^{(N_f,\alpha)}_N(y)
\label{PAP}
\eeq
(the factor 2 comes from the symmetricity of $A$).
Comparing eq. (\ref{PAP})  
with eq. (\ref{PCCP}) and (\ref{CC}) term by term we identify
\beqn
2 \bar{A}_i & =&   - a_N^2 \chi(\beta) I_i^{(N-1,N-1)} =
- a_N^2  \chi(\beta)
d_{N-1,i}^{(N_f,\alpha)}({\bf m})
P_{N-1}^{(N_f,\alpha)}(-m_i^2;{\bf m}) \ , \quad i=1,\ldots,N_f \nonumber \\
2 \bar{A}_{N_f+1} &=&  - \alpha a_N^2 \chi(\beta) I_0^{(N-1,N-1)}  =
  - \alpha a_N^2   \chi(\beta)
d_{N-1,N_f+1}^{(N_f+1,\alpha-1)}({\bf m},0) 
P_{N-1}^{(N_f,\alpha)}(0;{\bf m}) \ , \label{Adiag}
\eeqn
from which, the matrix $A$ is completely determined\footnote{
Notice that with our choice of $\psi$ functions, all the elements 
$\bar{A}_i$ are determined through the function $C(x)$ only.}.
Moreover, it is a result of Widom that the matrix $A$ is {\it uniquely}
 determined 
by the given choice of $\psi_i$ functions \cite{Widom}. 
Therefore it is garantueed
that all the lower-order terms in eq. (\ref{DK}) and eq. (\ref{Aij})
match exactly.\\
 To calculate the corrections, we need the antisymmetric matrix $B$ in eq. 
(\ref{Bij}). 
 Unfortunately enough, the matrix elements $B_{ij}$ (which are a total
 number of $2n(2n-1)/2$ distinct elements) are much more 
difficult to evaluate than $A_{ij}$, and therefore our final result 
 for the corrections to the unitary kernel is given by eqs.
 (\ref{SS1}), (\ref{SS4}), and (\ref{Bij}), 
with $\psi$'s  explicitly given in eqs. (\ref{psii1}), (\ref{psii21}) and
(\ref{psii22}) and the matrix elements of $A$ are given in (\ref{Adiag}).\\
Although such a final expression is not terribly simple, it is at least 
suitable for direct numerical evaluations. However,
 its major usefulness appears when one considers the microscopic
 limit of it. 
In fact, one should remember that the physical interest is in the 
microscopic limit, since it is in this limit that 
universal properties of $\chi$RMT appear. In Section \ref{sectML} 
 we will study the kernel $S^{(\beta)}_N(x,y)$ exactly in this limit.\\

Before doing that, let us mention that 
all the results of the last three sections can be generalized easily to the case
where the weight function is 
\beq
\label{weight_c}
w^{(N_f,\alpha,c)}(x)= \prod_{i=1}^{N_f} (x+m_i^2) \,  x^{\alpha} \,   e^{-cx}
\ , 
\eeq
i.e. where an additional parameter $c>0$ has been introduced in the 
exponential.
We do not need here to repeat all the calculations above with this new weight 
function, because  it is possible to write down 
the $c$-dependence for any quantity, explicitly, 
just using scaling arguments.
For instance, from the identity
\beqn
\lefteqn{1=\int_0^{\infty} dx 
\left[ P_j^{(N_f,\alpha)}(x;{\sqrt{c} \bf m}) \right]^2
x^{\alpha} \,  \prod_{i=1}^{N_f} (x+c m_i^2) \, e^{-x} } \nonumber \\
& = & c^{\alpha+N_f+1} \int_0^{\infty} dx 
\left[ P_j^{(N_f,\alpha)}(cx;{\sqrt{c} \bf m}) \right]^2
x^{\alpha} \,  \prod_{i=1}^{N_f} (x+m_i^2) \, e^{-cx} \, ,
\eeqn
we can read off the orthonormal polynomials w.r.t. the new weight function in 
eq. 
(\ref{weight_c}): 
\beq
P_j^{(N_f,\alpha,c)}(x;{\bf m})=c^{\frac{\alpha+N_f+1}{2}}
P_j^{(N_f,\alpha)}(c x;\sqrt{c} {\bf m}) \ .
\eeq
The same argument applies to all the other functions. In general, 
 a quantity $g_{c=1}(x;{\bf m})$ defined with the measure 
$w^{(N_f,\alpha)}(x) dx$ turns out to be related to the same 
quantity $g_{c}(x;{\bf m})$ defined with the new measure 
$w^{(N_f,\alpha,c)}(x) dx$ according to this formula:
\beq
g_{c}(x;{\bf m})=c^{\gamma} g_{c=1}(c x;\sqrt{c} {\bf m}) \, ,
\eeq
where $\gamma$ is a suitable scaling exponent. 
The following table shows the scaling factor for the most 
important quantities in this paper.
\vspace{0.5cm}
\\
\begin{center}
\begin{tabular}{||l|c||l|c||}
\hline
\hline
Quantity at $c=1$ &  $\gamma$ for $c \neq 1$ &
Quantity at $c=1$ &  $\gamma$ for $c \neq 1$ \\
\hline
\hline
$w^{(N_f,\alpha)}(x)$ &  $-N_f-\alpha$ &
$A(x)$ & $1$ \\
\hline
$P_j^{(N_f,\alpha)}(x;{\bf m})$  &  $(\alpha+N_f+1)/2$  &
$B(x)$ & $1$\\
\hline
$k_n^{(N_f,\alpha)}({\bf m})$ & $n+(\alpha+N_f+1)/2$  &
$C(x)$ & $1$ \\
\hline
$\det[\Lambda^{(N_f,\alpha)}_{n}(x)]$ & $(N_f+1)(\alpha+1)/2$ &
$I_i^{(N,j)}$ & $1$\\
\hline
$h_n^{(N_f,\alpha)}({\bf m})$  &  $N_f(\alpha+2)$ &
$\psi_i(x)$ & $3/2$ \\
\hline
$c_{n,i}^{(N_f,\alpha)}({\bf m})$ & $-(\alpha+N_f+1)/2$ &
$D_i,D^{(\perp)}_i$ & $0$\\
\hline
$d_{n,i}^{(N_f,\alpha)}({\bf m})$ & $(1-\alpha-N_f)/2$ &
$A_{ij} $ &  $-1$ \\
\hline
$\varphi_n(x)$  &  $1/2$ &
$B_{ij} $ & $1$ \\
\hline
$U(x,z)$ & $2$ &
$\widehat{\varepsilon} \psi$ & $1/2$\\
\hline
$a_N({\bf m}) $ & $-1$ & 
$K_N^{(2)}(x,y)$ & $1$\\
\hline
\hline
\end{tabular}
\end{center}

\vspace{1cm}
\noindent
The only exception of this scaling rule, is given by the potential 
$V(x)$ which, for $c \neq 1$ is, $V(x)=c-\alpha \log x- \sum_{k=1}^{N_f} 
(x+m_k^2)$. \\
We conclude this Section remembering once again that at the very end one
 has to perform an important substitution. Namely, it is 
a prescription of this method to replace $N\rightarrow 2N$ when $\beta=4$ and 
 $w \to w^{2}$ for $\beta=1$. The latter substitution leads one to
consider double degenerate masses $m_{i}$, which means $N_f \to 2 N_f$, 
 and to take $\alpha \rightarrow 2\alpha$,  $c \rightarrow 2c$. We summarize 
these rules here:
\beq
\label{rules}
\begin{array}{llll}
\beta = 1 \quad : &  N_f \to 2 N_f,  & \alpha \to 2\alpha \ , & c \to 2c \\
\beta = 4 \quad : &  N\rightarrow 2N &  & \\
\end{array}
\ .
\eeq

\sect{The microscopic limit} 
\label{sectML}

The physical interest in studying RMT relies on its universal
properties which appear in the large-$N$ limit. In particular, we 
consider the double-microscopic limit of the scalar kernel:
\beq
\label{SS_ML}
\tilde{S}^{(\beta)}(\zeta_1,\zeta_2)=\lim_{N \to \infty}
\frac{c}{N^2}S_{N}^{(\beta)}(\frac{c \zeta_1}{N^2},
\frac{c \zeta_2}{N^2}) \  , \quad m_i=\frac{\mu_i \sqrt{c}}{N} \ , \ 
c=\frac{\beta N}{2} \ , 
\eeq
where $\zeta_1,\zeta_2$, and $\mu_i$ are kept fixed. In 
 general, the microscopic limit of a quantity 
$g_{c}(x,{\bf m})$ in this paper is:
\beq
\label{g_ML}
\tilde{g}_c(\zeta,{\bf \mu})=
\lim_{N \to \infty} N^{\delta} c^{\gamma} 
g_{c=1}(\frac{c \, \zeta}{N^2}, \frac{\bmu\sqrt{c}}{N})= 
\left( \frac{\beta}{2} \right)^{\gamma} 
\lim_{N \to \infty} N^{\delta+\gamma} 
g_{c=1}(\frac{\beta \zeta}{2 N},\sqrt{\frac{\beta}{2N}}
 {\bmu })
\eeq
where the exponent $\gamma$ can be read off from the table at the end 
of Section $6$ and 
$\delta$ is a suitable exponent which is chosen 
such that a finite non-zero limit exists. According to eq.~(\ref{g_ML}) we may
 obtain the microscopic limit of $g_{c}(x,{\bf m})$ simply by computing
 the microscopic limit of $g_{c=1}(x,{\bf m})$ with the scaling $x=\zeta/N$,
 ${\bf m}=\bmu/\sqrt{N}$, and then evaluating the result at the points 
$\zeta \to \tilde{\zeta} \equiv \zeta \beta/2, \  \bmu \to \tilde{\bmu} \equiv \sqrt{\beta/2} \, \bmu$. Following this strategy 
 when computing the microscopic limit of $S^{(\beta)}(x,y)$ 
in eq. (\ref{SS_ML}), we 
consider $c=1$ in the
 rest of this section, and then at the very end we substitute  
$\zeta \to \tilde{\zeta}$ and $\bmu \to \tilde{\bmu}$.\\ 
In eq. (\ref{SS1}) and (\ref{SS4}) the scalar kernel 
$S_N^{(\beta)}(x,y)$ is expressed 
 in terms of the unitary kernel $K^{(2)}_N(x,y)$ and the $\psi_i$ functions which
all are determined explicitly in Section \ref{sect_hilbert}. 
Hence we can now compute the microscopic 
large-$N$ limit of the scalar kernel $S_N^{(\beta)}(x,y)$.
 The basic ingredient is the set of the  
 orthonormal polynomials $P_n^{(N_f,\alpha)}$ and   
their microscopic limit. Such a limit is straightforward
 analytically. In fact, the polynomials are defined completely in terms of
 determinants of matrices of the form as in eq. (\ref{Lambda}).
If one naively takes  the microscopic limit of such determinants, then
 it eventually ends up with an indeterminate form. The situation 
 is actually analogous to the case of degenerate masses, i.e. one has to  
substitute rows or columns in the determinants with 
suitable linear combinations before taking the microscopic limit. 
For instance, the generic
 determinant can conveniently be written as (we use the shortened notation 
 $t_0=x, t_i=-m_i^2$):
\[
\det[\Lambda_N^{(N_f,\alpha)}]=
\det_{il} [ P^{(0,\alpha)}_{N+l}(t_i) ] =
 \frac{\det_{il} [ L^{(\alpha)}_{N+l}(t_i) ]}{\prod_{k=N}^{N+N_f}
\sqrt{h_{k}^{\alpha}} }
= \frac{\det_{il} [ t_i^{l}L^{(\alpha+l)}_N(t_i) ]}{\prod_{k=N}^{N+N_f}
\sqrt{h_{k}^{\alpha}} \prod_{p=1}^{N_f}[-(N+p)]^{N_f-p+1}} \ , 
\]
with $i,l=0,\ldots, N_f$. In the last 
equality we iteratively used the recurrence relation 
$x L^{(\alpha+1)}_n(x)=(n+\alpha+1) L^{(\alpha)}_n(x)-(n+1)
L^{(\alpha)}_{n+1}
(x)$.  Now substituting the microscopic scaling 
 $t_i \to z_i/N$, one obtains the following large-$N$ behaviour  
\beq
\det[\Lambda_N^{(N_f,\alpha)}] 
\sim
(-1)^{\frac{N_f (N_f+1)}{2}}
N^{\frac{(N_f+1)(\alpha-N_f)}{2}} \det_{il} [ z_i^{\frac{l-\alpha}{2}}
J_{\alpha+l}(2 \sqrt{z_i}))] \ , 
\label{finalML}
\eeq
since $\lim_{N \to \infty} L_N^{(\alpha)}(x/N)/N^{\alpha} =J_{\alpha}(2 
\sqrt{x})/x^{\alpha/2}$. 
When the argument of the Bessel functions is complex (i.e., when 
$t_i=-m^2_i$), one will make use of  $J_{\nu}(i z)=i^{\nu} I_{\nu}(z)$.
Finally, from the previous simple arguments it follows that the
microscopic
 limit of all the quantities evaluated in this paper, can be
obtained using the determinant in eq. (\ref{finalML}) instead of the
determinant at finite-$N$ given in eq.~(\ref{Lambda}).\\
Let us look at some explicit examples. For sake of simplicity, 
 we consider $N_f$ non-degenerate masses first.
The microscopic limit of the 
normalization factor $h^{(N_f,\alpha)}_N$ in eq.~(\ref{norm}) is: 
\beq
\label{h_ML}
\tilde{h}^{(N_f,\alpha)}({\bmu}) =
[\Delta^{(\alpha)}(\bmu)]^2 \ , 
\eeq
where $\bmu \equiv \{\mu_1,\mu_2, \ldots, \mu_{N_{f}} \}$ and
\beq
\Delta^{(\alpha)}(\mu_1,\mu_2, \ldots, \mu_{N_{f}}) \equiv
\det_{pq} [ \mu_p^{q-\alpha} I_{\alpha+q}(
2 \mu_p)] \ , 
\eeq
where $p=1,\ldots,N_{f}$, $q=0,\ldots,N_{f}-1$. In this definition, we 
 assume that the range of $p$ and $q$ is determined by 
the number of arguments of $\Delta^{(\alpha)}$. 
We can also compute the microscopic limit of the 
highest coefficient $k^{(N_f,\alpha)}_N({\bf m})$ 
in an analogous way:
\beq
\label{k_ML}
\tilde{k}^{(N_f,\alpha)}({\bmu}) = 1 \ ,
\eeq
because the ratio of the two determinants in eq.~(\ref{knnf})
 equals unity in the microscopic large-$N$ limit.  
From eq.~(\ref{CT}), (\ref{k_ML}) and (\ref{h_ML}) 
we immediately read off the microscopic limit
 of the orthonormal polynomials for $N_f$ massive fermions, that is:
\beq
\label{P_ML}
\tilde{P}^{(N_f,\alpha)}(\zeta;\bmu) =
\frac{\Delta^{(\alpha)}_1(\zeta,\bmu)}
{\Delta^{(\alpha)}(\bmu) 
\prod_{i=1}^{N_f} (\zeta+\mu_i^2)} \ .
\eeq
where $\Delta_1^{(\alpha)}(\zeta,\bmu)$ is 
\beq
\label{delta_prime}
\Delta_1^{(\alpha)}(\zeta,\bmu) \equiv
\det 
\left(
\begin{array}{cccc}
\zeta^{\frac{-\alpha}{2}} J_{\alpha}(2 \sqrt{\zeta}) & 
-\zeta^{\frac{1-\alpha}{2}} J_{\alpha+1}(2 \sqrt{\zeta}) &
\cdots &
(-1)^{N_f} \zeta^{\frac{N_f-\alpha}{2}} J_{\alpha+N_f}(2 \sqrt{\zeta}) \\
\mu_1^{-\alpha} I_{\alpha}(2 \mu_1) & 
\mu_1^{1-\alpha} I_{\alpha+1}(2 \mu_1) &
\cdots &
\mu_1^{N_f-\alpha} I_{\alpha+N_f}(2 \mu_1) \\
\cdots & \cdots & \ddots & \cdots  \\
\mu_{N_f}^{-\alpha} I_{\alpha}(2 \mu_{N_f}) & 
\mu_{N_f}^{1-\alpha} I_{\alpha+1}(2 \mu_{N_f}) &
\cdots &
\mu_{N_f}^{N_f-\alpha} I_{\alpha+N_f}(2 \mu_{N_f}) 
\end{array}
\right) \, .
\eeq
The microscopic limit of the 
coefficients $d_{N,i}^{(N_f,\alpha)}({\bf m})$
in eq.~(\ref{dcoeff_1}) is
\beq
\tilde{d}_{i}^{(N_f,\alpha)}(\bmu) 
=
\frac{\Delta^{(\alpha)}(\bmu_{\neq i}) \prod_{j \neq i}^{N_f} (\mu_j^2-\mu_i^2)}
{(-1)^{i+1}\Delta^{(\alpha)}(\bmu)} \, .
\eeq
The microscopic limit of the weight $w^{(N_f,\alpha)}(x)$ is given by
\beq
\tilde{w}(\zeta)=
\zeta^{\alpha} \prod_{i=1}^{N_f} (\zeta+\mu_i^2) \ , 
\eeq
which appear in the following expression for the microscopic 
unitary kernel (from eq.~(\ref{Kcompact})):
\beqn
\tilde{K}^{(2)}(\zeta_1,\zeta_2) &=&
\frac{ \tilde{P}^{(N_f+1,\alpha)}(\zeta_1;{\bmu},\sqrt{-\zeta_2}) }
{\tilde{d}_{N_f+1}^{(N_f+1,\alpha)}({\bmu},\sqrt{-\zeta_2}) } 
\sqrt{\tilde{w}(\zeta_1) \tilde{w}(\zeta_2)}
\nonumber \\
&=&
\sqrt{\zeta_1^{\alpha}\zeta_2^{\alpha}}
\frac{(-1)^{N_f}\Delta_1^{(\alpha)}(\zeta_1;\bmu,\sqrt{-\zeta_2})}
{\Delta^{\alpha}(\bmu) (\zeta_1-\zeta_2) \sqrt{\prod_{j=1}^{N_f} 
(\zeta_1+\mu_j^2) (\zeta_2+\mu_j^2)}} \, . \label{K_ML}
\eeqn
The microscopic limit of the $\psi_i$ functions is:
\beqn
\tilde{\psi}_i(\zeta) &=&
\frac{\Delta^{(\alpha)}_1(\zeta,\bmu,\sigma_i)}
{\Delta^{(\alpha)}(\bmu,\sigma_i) 
(\zeta+\sigma_i^2) \sqrt{\prod_{j=1}^{N_f} (\zeta+\mu_j^2)}} 
\sqrt{\zeta^{\alpha}} \ , \quad i=1,\ldots,N_f+1 \ , \nonumber \\
\tilde{\psi}_{N_f+1+j}(\zeta) 
&=& 
\frac{\Delta^{(\alpha)}_1(\zeta,\bmu_{\neq j})}
{\Delta^{(\alpha)}(\bmu_{\neq j}) 
\sqrt{\prod_{j=1}^{N_f} (\zeta+\mu_j^2)}} 
\sqrt{\zeta^{\alpha}} \ , \quad j=1,\ldots,N_f \ , \nonumber \\
\tilde{\psi}_{2N_f+2}(\zeta) 
&=&
\frac{\Delta^{(\alpha-1)}_1(\zeta,\bmu)}
{\Delta^{(\alpha-1)}(\bmu) 
\sqrt{\prod_{j=1}^{N_f} (\zeta+\mu_j^2)}} 
\sqrt{\zeta^{\alpha-2}} \ , \nonumber 
\eeqn
where $\sigma_i \equiv m_i$ for $i=1,\ldots, N_f$ and $\sigma_{N_f+1} \equiv 0$. The microscopic limit of the diagonal elements of the matrix 
$\bar{A}$ is (from eq.~(\ref{Adiag})):
\beqn
2 \tilde{A}_i &=&  
-\chi(\beta)
\tilde{d}_{i}^{(N_f,\alpha)}({\bmu})
\tilde{P}^{(N_f,\alpha)}(-\mu_i^2;{\bmu}) 
\ , \quad  \  i=1,\ldots,N_f \label{A1_ML}  \ , \\
2 \tilde{A}_{N_f+1} &=& 
-\alpha \chi(\beta)
\tilde{d}_{N_f+1}^{(N_f+1,\alpha-1)}({\bmu},0) 
\tilde{P}^{(N_f,\alpha)}(0;{\bmu}) \nonumber \\
&=&
-\alpha \chi(\beta) 
\frac{\Delta^{(\alpha-1)}(\bmu)}{\Delta^{(\alpha)}(\bmu) } 
\frac{\Delta_1^{(\alpha)}(0;\bmu)}{\Delta^{(\alpha-1)}(0,\bmu) }
\, . \label{A2_ML} 
\eeqn
In equation (\ref{A1_ML}) the orthogonal polynomials $\tilde{P}^{(N_f,\alpha)}$ are evaluated at $\zeta=-\mu_i^2$. As we have 
already discussed for the degenerate
 massive case, this situation 
should be considered in a limiting sense of eq.~(\ref{P_ML}), that is $\zeta \to -\mu_i^2$. 
In particular,  in the ratio $\Delta_1^{(\alpha)}(\zeta,\bmu) / 
\prod_{i=1}^{N_f} (\zeta+\mu_i^2)$ one substitutes  each element of 
the first line of the matrix in eq.~(\ref{delta_prime}) 
with its derivative with respect to $\zeta$, evaluated at 
 $\zeta=-\mu_i^2$.\\ 
Finally, the elements of the matrix $B$ and the terms $\hat{\varepsilon} \psi_i$ 
are given in eq.~(\ref{Bij}) and eq.~(\ref{epspsi}), respectively. Let 
$\tilde{B}_{ij}$ and $\widetilde{\varepsilon \psi}_i$ be their microscopic 
large-$N$ limit, respectively. By means of eq.~(\ref{Cmatrix}), one obtains
also $\tilde{C}$, $\tilde{C}_{00}$ and $\tilde{C}_0$. 
Putting together all of the terms above, 
we obtain the microscopic
limit of the scalar kernel $S_{N}^{(\beta)}(x,y)$. 
It is very important at this point to note that the number of corrections 
to the unitary kernel $K^{(2)}_N(x,y)$ appearing in $S_{N}^{(\beta)}(x,y)$
is indeed independent of $N$,  because 
it depends only on the potential $V(x)$, and 
thus in the double-microscopic large-$N$ limit the 
scalar kernel $S^{(\beta)}_N(x,y)$ is still obtained through  the
formulas (\ref{SS1}), (\ref{SS4}). Namely, in 
the orthogonal case $\beta=1$, the scalar kernel is:
\beq
\label{S1_M}
\tilde{S}^{(1)}(\zeta_1,\zeta_2) = 
\tilde{K}^{(2)}(\zeta_1,\zeta_2)-\sum_{i\leq n,j=1}^{2n} 
[\tilde{A}\tilde{C}(I-\tilde{B}\tilde{A}\tilde{C})^{-1}]_{ji}
\tilde{\psi}_{i}(\zeta_1)\widetilde{\varepsilon\psi}_{j}(\zeta_2) \ . 
\eeq
In fact, the required substitution $c \to 2c$, which effectively means $\beta \to 2 \beta$ in this case, is compensated by the $\beta$-scaling required for recovering
the general case $c \neq 1$, as discussed at the beginning of this section.
Moreover we have to put $\alpha \to 2\alpha$ and $N_f \to 2 N_f$, the latter 
meaning that all the masses are double degenerate. 
Indeed the degenerate-masses case does not present any peculiar problem, 
and by applying the 
same technique as described after eq.~(\ref{knnf}), the only consequence is
 that one has to consider
higher-order derivatives of Bessel functions in all the determinants.\\
In the symplectic case $\beta=4$,  one has to apply the 
substitution $N \to 2N$. Accordingly, the microscopic limit of the unitary 
kernel reads:
\beq
\lim_{N \to \infty} 
N^{\delta} K^{(2)}_{2N}(\frac{
\zeta_1}{2N},\frac{
\zeta_2}{2N})=
\tilde{K}^{(2)}(2 \zeta_1,2 \zeta_2) \, ,
\eeq
and the scalar kernel is 
\beq
\label{S4_M}
\tilde{S}^{(4)}(\zeta_1,\zeta_2)  =  
\left[
2\tilde{K}^{(2)}(2 \tilde{\zeta_1},2 \tilde{\zeta_2})-2\sum_{i>n,j=1}^{2 
n}[\tilde{A}_{0}\tilde{C}_{00}^{-1}\tilde{C}_{0}]_{ij} \ 
\widetilde{\psi}_{i}(2 \zeta_1) \ \widetilde{\varepsilon\psi}_{j}(2 \zeta_2)
\right]_{
\bmu \to  \sqrt{2} \bmu  
} \ ,
\eeq
where once again the final  $\beta$-scaling is for recovering
 $c=N \beta/2 \neq 1$. The only elements in the previous formulas which 
require further attention are the large-$N$ quantities 
$\tilde{B}_{ij}$ and $\widetilde{\varepsilon\psi}_j$. 
In order to obtain them, we need to evaluate the large-$N$ behaviour of 
the integrals defining the matrix elements $B_{ij}$ and $\varepsilon \psi_i$. 
When doing that, it could be advantageous to exchange the integrals with 
the large-$N$ microscopic limit, i.e. substituting the microscopic 
expansion of the functions $\psi_i$ into the integrals. Such an  
interchange can be done under suitable smoothness assumptions.    
In \cite{sen98v} this issue has been investigated carefully for the 
massless case. It turns out 
that for $\beta=4$ the procedure holds without any problems, 
whereas for $\beta=1$ one has to consider also an additional 
contribution coming from the soft-edge of the spectrum. 
Since the case of one mass taken to zero or  to infinity in 
our formulas reproduces results with one additional massless flavor
and one massless flavor less, respectively, 
we shall assume that the interchanging procedure, when $\beta=4$, 
is permitted in the intermediate mass region as well. 
Thus, in the $\beta=4$ case we can write
\beqn
\tilde{B}_{ij} &=&
\int_{0}^{\infty} \int_{0}^{\infty} dx dy \,  \varepsilon(x-y) \,
 \tilde{\psi}_i(x) \,  \tilde{\psi}_j(y) \ ,  \\
\widetilde{\varepsilon \psi}_{j}(y) &=& 
\int_{0}^{\infty} dx  \varepsilon(x-y) \tilde{\psi}_j(x) \  ,  
\eeqn
with $i,j=1,\ldots,2N_f+2$. 
The question whether such a simplification is allowed 
 for $\beta=1$ case, can be addressed by means of numerical 
investigations. Nevertheless, with or without exchanging the
 integrals and limits,
 from the very general result in eq.~(\ref{S1_M}) and 
eq.~(\ref{S4_M}) one obtains analytic expressions for the 
microscopic limit of all the correlation functions of the 
general massive chiral ensembles for $\beta=1$ and $\beta=4$. 
For instance, the microscopic 
spectral density is the simplest one and it is given by 
$\rho^{(\beta)}(x)=\tilde{S}^{(\beta)}(x,x)$. 
Switching to the real eigenvalues $\xi_i$ of the Dirac operator,
 by means of $\xi^2=x$, one obtains the microscopic 
spectral density of the Dirac operator $\rho^{(\beta)}_{\mathcal D}(\xi)=
2 |\xi| \rho^{(\beta)}(\xi^2)$.\footnote{In general, the $k$-point correlation function $\rho^{(\beta)}_{\mathcal D}(\xi_1,\ldots,\xi_k)$ of the spectrum of the Dirac operator ${\mathcal D}$ is given by the change of variables $\xi_j^2=\lambda_j$ in 
eq.~(\ref{corr}).} If universality arguments apply to correlation  functions 
of the general massive Chiral Ensembles, 
then the expressions obtained from eq.~(\ref{S1_M}) and 
eq.~(\ref{S4_M}), through eq.~(\ref{corr}), are the 
very natural candidates.\\

\sect{Conclusions}   

In this paper we considered the problem of computing the 
 correlation functions for massive Dirac spectra of four-dimensional QCD. 
Starting from the fact that RMT is just a simple and effective tool for 
 calculating actual spectral correlations in the infrared regime 
(which in principle also can be obtained from direct calculations in 
terms of finite-volume partition function as in \cite{PoGer} ) we 
consider a matrix model with matrices either in the $\chi$GOE 
or in the $\chi$GSE. The former corresponds to four-dimensional QCD 
with $N_f$ fermions in the fundamental representation and $SU(2)$ gauge theory, 
 the latter corresponds to four-dimensional QCD with $N_f$ fermions in the 
adjoint representation and $SU(N_c\geq 2)$ gauge theory. Matrix models 
 with orthogonal and symplectic matrices naturally leads to the application 
of skew-orthogonal polynomials, which have the drawback that 
at present they are difficult to 
 determine and handle in the microscopic 
 large-$N$ limit. Therefore we investigated whether 
the recent technique proposed by H.~Widom for dealing with OE and SE 
 using standard orthogonal polynomials of the UE, could be applied effectively 
in
 our actual case. We have succeeded in deriving the orthonormal polynomials 
 for the general massive fermion case, in an explicit and closed form. 
 We found explicit formulas for the $\psi$ functions which are the basic
 ingredients for computing the scalar kernels $S^{(\beta)}_N(x,y)$ for massive 
fermions, 
 when $\beta=1$ and $\beta=4$. In particular, such scalar 
 kernels are expressed as the unitary kernel $K^{(2)}_N(x,y)$ plus
 a finite number of corrections, which depend only on one-dimensional
 and two-dimensional integrals involving the functions $\psi_i$. Cases with degenerate masses, 
 or massless fermions can then be obtained just as simple limits of the formulas 
 we have derived. We obtained fully analytical formulas for the scalar kernels 
$\tilde{S}^{(1)}$ and $\tilde{S}^{(4)}$ in the double 
microscopic large-$N$ limit. From such formulas one can in principle derive 
all the microscopic correlation functions of the Chiral Orthogonal and 
Chiral Symplectic Ensembles, with an arbitrary number of flavors, 
arbitrary masses and arbitrary topological charge.\\ 
We emphasize that the issue of universality has not been considered in the present paper. However, the  fact that the method of Widom can be applied to  
 four-dimensional QCD with massive fermions is indeed encouraging 
 and provides a new general framework where it might be possible to analyze 
 some of the still open issues. We believe that 
the  present framework is the most suitable setting for proving universality
 of these massive cases. 
\vspace{1cm}

\indent
\underline{Acknowledgements}: 
We wish to thank  P.~H.~Damgaard for continuous encouragements, 
several useful discussions and for carefully reading the manuscript.  
The work of G.~V.~ was supported in part by The Niels Bohr Institute 
of Copenhagen and in part by the EU grant no. HPRN-CT-1999-00161.\\

\begin{appendix}

\sect{Normalization factors}\label{app_norm}
The orthonormality condition of polynomials is an essential requirement
for the proper use of Widom's method. Therefore, an explicit expression 
for the normalization factors of the orthogonal polynomials for the 
general $N_f$-flavor case is of primary importance. Even though it is a 
classical
 result,  
we choose to present in this appendix a derivation of the normalization factors 
$h^{(N_f,\alpha)}_n$, 
and of the highest coefficients $k^{(N_f,\alpha)}_n$, i.e. eq. (\ref{norm}) 
and eq. (\ref{knnf}), respectively.  
Substituting Christoffel's formula eq. (\ref{CT}) into the orthonormality 
condition eq. (\ref{orthonorm}) evaluated at $i=j=n$, yields:
\beqn
\sqrt{h^{(N_f,\alpha)}_n({\bf m})} & = & \int_0^{+\infty} 
dx \,  x^{\alpha} e^{-x} 
P_n^{(N_f,\alpha)}(x;{\bf m})
\det \left[ \Lambda^{(N_f,\alpha)}_{n}(x) \right]  \nonumber \\
 & = & 
\int_0^{+\infty} dx \,  x^{\alpha} e^{-x} 
P_n^{(N_f,\alpha)}(x;{\bf m}) 
\sum_{k=0}^{N_f} 
P_{n+k}^{(0,\alpha)}(x) (-1)^k 
\det \left[ \Lambda^{(N_f,\alpha)}_{n,k+1} \right] \, ,\label{laplace}
\eeqn
where we in the second equation used the so-called Laplace expansion
 of the determinant of a matrix with respect to its first row. Since 
$\int_0^{+\infty} dx \,  x^{\alpha} e^{-x} 
x^n  P_{j}^{(0,\alpha)}(x)=\delta_{nj}/k_n^{(0,\alpha)}$, for $n \leq j$, 
one has that the only non-vanishing term in the sum in eq. (\ref{laplace}) 
is when $k=0$. Therefore we obtain: 
\beq
\label{norm_first}
\sqrt{h^{(N_f,\alpha)}_n({\bf m})}=\frac{k_n^{(N_f,\alpha)}({\bf 
m})}{k_n^{(0,\alpha)}} 
\det \left[ \Lambda^{(N_f,\alpha)}_{n,1}  \right] \ ,
\eeq
with the coefficient $k_n^{(0,\alpha)}$ given in eq. (\ref{high_lagrre}).
The still unknown highest coefficients $k_n^{(N_f,\alpha)}$ can 
 easily be determined as follows:
\beqn
k_n^{(N_f,\alpha)}({\bf m}) & = & \lim_{x \rightarrow +\infty} 
\frac{P_n^{(N_f,\alpha)}(x;{\bf m})}{x^n}=\lim_{x \rightarrow +\infty} 
\frac{\sum_{k=0}^{N_f} P_{n+k}^{(0,\alpha)}(x) (-1)^k
\det \left[ \Lambda^{(N_f,\alpha)}_{n,k+1} \right]}
{x^n \sqrt{h^{(N_f,\alpha)}_n({\bf m})}  \prod_{i=1}^{N_f} (x+m_i^2)} \nonumber 
\\
 &=& 
\frac{k_{n+N_f}^{(0,\alpha)} (-1)^{N_f}
\det \left[ \Lambda^{(N_f,\alpha)}_{n,N_f+1} \right]}
{\sqrt{h^{(N_f,\alpha)}_n({\bf m})}} \label{quasi} \ . 
\eeqn
In the first line we used Christoffel's formula eq. (\ref{CT}) and the Laplace 
expansion of the determinant, whereas the second line comes from the 
observation that only the last term in the sum contributes in the 
 large-$x$ limit.
Combining eq. (\ref{quasi}) with eq. (\ref{norm_first}) one finally has
\beqn
h^{(N_f,\alpha)}_n ({\bf m})&=& (-1)^{N_f}
\frac{k_{n+N_f}^{(0,\alpha)}}{k_n^{(0,\alpha)}} 
\det \left[ \Lambda^{(N_f,\alpha)}_{n,1}  \right]
\det \left[ \Lambda^{(N_f,\alpha)}_{n,N_f+1}  \right] \ , \\
k^{(N_f,\alpha)}_n ({\bf m}) &=& \sqrt{  (-1)^{N_f}
 k_n^{(0,\alpha)}  k_{n+N_f}^{(0,\alpha)}
\frac{\det \left[ \Lambda^{(N_f,\alpha)}_{n,N_f+1}  \right]}
{\det \left[ \Lambda^{(N_f,\alpha)}_{n,1}  \right] }  }  \ , 
\eeqn
where the coefficients $k^{(0,\alpha)}_n$ are given in eq. 
(\ref{high_lagrre}).

\sect{Some integrals} \label{app_int}
In this appendix we evaluate the integrals in  
eq. (\ref{IN0}) and eq. (\ref{INj}), that is:
\beq
\label{First_appB}
\left\{
\begin{array}{lcl}
I^{(N,j)}_0 & = & \langle 
P_N^{(N_f,\alpha)}(z;{\bf m}),\frac{P_j^{(N_f,\alpha)}(z;{\bf m})}{z} 
\rangle_{N_f,\alpha} \\
& & \\
I^{(N,j)}_i & = &  \langle 
P_N^{(N_f,\alpha)}(z;{\bf m}),\frac{P_j^{(N_f,\alpha)}(z;{\bf m})}{z+m_i^2} 
 \rangle_{N_f,\alpha} \quad {\rm for} \ i=1,2,\ldots,N_f 
\end{array}
\right.   \ .
\eeq
Using eq. (\ref{diag_move}) (i.e. adding $m_{N_f+1}=0$) the integral 
$I^{(N,j)}_0$ can also be written as
\beq 
I^{(N,j)}_0 = \langle 
P_N^{(N_f+1,\alpha-1)}(z;{\bf m},0),
\frac{P_j^{(N_f+1,\alpha-1)}(z;{\bf m},0)}{z} 
\rangle_{N_f+1,\alpha-1} \ , 
\eeq
therefore it can be obtained from $I^{(N,j)}_i$ in eq.~(\ref{First_appB}), by 
applying the substitutions  
$N_f \to N_f+1$, $\alpha \to \alpha-1$,  
$\{ {\bf m} \} \to \{ {\bf m},0 \}$ and $i \to N_f+1$.\\
The integral $I^{(N,j)}_i$ can be evaluated as follows. 
First we apply eq. (\ref{c-darboux}) and then we 
use the orthonormality condition (\ref{orthonorm}), that is
\beqn
I^{(N,j)}_i &=& 
d_{N,i}^{(N_f,\alpha)}({\bf m})
d_{j,i}^{(N_f,\alpha)}({\bf m}) 
\sum_{r=0}^{N} 
\sum_{s=0}^{j}
P_r^{(N_f-1,\alpha)}(-m_i^2;{\bf m}_{\neq i}) \nonumber \\
& \times & P_s^{(N_f-1,\alpha)}(-m_i^2;{\bf m}_{\neq i})
\langle
P_r^{(N_f-1,\alpha)}(z;{\bf m}_{\neq i}),
P_s^{(N_f-1,\alpha)}(z;{\bf m}_{\neq i})
\rangle_{N_f-1,\alpha}  \nonumber \\
&=&
d_{N,i}^{(N_f,\alpha)}({\bf m})
d_{j,i}^{(N_f,\alpha)}({\bf m}) 
\sum_{r=0}^{\min(N,j)}
\left[ P_r^{(N_f-1,\alpha)}(-m_i^2;{\bf m}_{\neq i}) \right]^2 \, . 
\eeqn
Using eq. (\ref{c-darboux}) again,  we finally end up with 
\beqn
\label{I1}
I^{(N,j)}_i & = & \frac{d_{N,i}^{(N_f,\alpha)}({\bf m})
d_{j,i}^{(N_f,\alpha)}({\bf m}) }
{d_{\min(N,j),i}^{(N_f,\alpha)}({\bf m})}
P_{\min(N,j)}^{(N_f,\alpha)}(-m_i^2;{\bf m}) \nonumber \\
& = & d_{\max(N,j),i}^{(N_f,\alpha)}({\bf m})
P_{\min(N,j)}^{(N_f,\alpha)}(-m_i^2;{\bf m})\, .
\eeqn
Therefore the integral $I^{(N,j)}_0$ is 
\beqn
\label{I0}
I^{(N,j)}_0 & = & d_{\max(N,j),N_f+1}^{(N_f+1,\alpha-1)}({\bf m},0)
P_{\min(N,j)}^{(N_f+1,\alpha-1)}(0;{\bf m},0) \nonumber \\
&=& d_{\max(N,j),N_f+1}^{(N_f+1,\alpha-1)}({\bf m},0)
P_{\min(N,j)}^{(N_f,\alpha)}(0;{\bf m}) \ .  
\eeqn
A final comment: one could wonder about the meaning of
 the ``degenerate'' expressions 
$P_n^{(N_f+1,\alpha)}(0;{\bf m},0)$ and 
$P_n^{(N_f,\alpha)}(-m_i^2;{\bf m})$. But as in the case of
 degenerate masses, they should be understood in the 
limit sense $x \to 0$ or $x \to -m_i^2$, respectively. 
Such a limit is easily evaluated from eq. (\ref{CT}) by 
subtracting the $i$-th row from the first row of 
the matrix $\Lambda^{(N_f,\alpha)}_n$ (eq. (\ref{Lambda})), and dividing it by 
the term $(x+m_i^2)$ stemming from the product in the denominator
of eq. (\ref{CT}). Therefore, the limit $x \rightarrow -m_i^2$  of
eq. (\ref{CT}) is equivalent to substituting all the polynomials
$P^{(0,\alpha)}_j(x)$ in the first row of $\Lambda^{(N_f,\alpha)}_n$, 
with the corresponding 
derivatives $P^{(0,\alpha)}_j{\,}'(x)$ evaluated at $x=-m_i^2$.

\sect{The functions $\psi_j(x)$} \label{app_Dcoef}

In this appendix we find a compact and useful expression 
for the functions $\psi$. We will show the remarkable fact that 
 every $\psi$ function is proportional to a polynomial
 $P^{(N_f,\alpha)}_n$. Let us first consider the functions in 
 the Hilbert space ${\mathcal H}$. Substituting the 
coefficients $D_i$ eq. (\ref{D}) into the linear combination
 eq. (\ref{psi}) we obtain:
\beqn
\psi_i(x)  & = & \frac{P_{N-1}^{(N_f,\alpha)}(-\sigma^2_i;{\bf m})
P_N^{(N_f,\alpha)}(x;{\bf m})-
P_N^{(N_f,\alpha)}(-\sigma_i^2;{\bf m})
P_{N-1}^{(N_f,\alpha)}(x;{\bf m})
}{(x+\sigma_i^2)P_{N-1}^{(N_f,\alpha)}(-\sigma_i^2;{\bf m})} 
\sqrt{w^{(N_f,\alpha)}(x)} \nonumber \\
& = &  - \frac{P_{N-1}^{(N_f+1,\alpha)}(x;{\bf m},\sigma_i)}
{P_{N-1}^{(N_f,\alpha)}(-\sigma^2_i;{\bf m}) \, 
c^{(N_f+1,\alpha)}_{N-1,N_f+1}({\bf m},\sigma_i)} \sqrt{w^{(N_f,\alpha)}(x)} \ ,
\nonumber \\
\eeqn
where $\sigma_i \equiv m_i$ for $i=1,\ldots, N_f$, $\sigma_{N_f+1} 
\equiv 0$ and in the second line we used the formula (\ref{CT_recurr}).  
From eq. (\ref{CT_coeff}) we obtain:
\beq
\psi_i(x)=k_N^{(N_f,\alpha)}({\bf m})
\frac{P_{N-1}^{(N_f+1,\alpha)}(x;{\bf m},\sigma_i)}
{k_{N-1}^{(N_f+1,\alpha)}({\bf m},\sigma_i)} \sqrt{w^{(N_f,\alpha)}(x)}
\ , \quad i=1,\ldots,N_f+1 \, .
\eeq
Also the $\psi$ functions in the orthogonal Hilbert space 
${\mathcal H}^{\perp}$ can be written in a simplified form. 
In fact, let us first 
notice that by isolating the last term of the sum in eq. (\ref{c-darboux}) 
and using again eq. (\ref{c-darboux}) on the remaining sum, one has:
\beq
P_N^{(N_f,\alpha)}(x;{\bf m})  =   
d_{N,i}^{(N_f,\alpha)}({\bf m}) 
\left(
\frac{ P_{N-1}^{(N_f,\alpha)}(x;{\bf m})}
{d_{N-1,i}^{(N_f,\alpha)}({\bf m})} +
P_N^{(N_f-1,\alpha)}(-m_i^2;{\bf m}_{\neq i})
P_N^{(N_f-1,\alpha)}(x;{\bf m}_{\neq i}) 
\right) \ , \nonumber 
\eeq
that is 
\beq
P_N^{(N_f,\alpha)}(x;{\bf m}) -
\frac{d_{N,i}^{(N_f,\alpha)}({\bf m})}{d_{N-1,i}^{(N_f,\alpha)}({\bf m})}
P_{N-1}^{(N_f,\alpha)}(x;{\bf m})=   
d_{N,i}^{(N_f,\alpha)}({\bf m})  P_N^{(N_f-1,\alpha)}(-m_i^2;{\bf m}_{\neq i})
P_N^{(N_f-1,\alpha)}(x;{\bf m}_{\neq i}) \ . 
\eeq
From eq. (\ref{Dperp}) and (\ref{dperpcoeff}) (or (\ref{dperpcoeffbis})), 
we exactly obtain the l.h.s. of last formula, therefore 
\beqn
\psi_{N_f+1+i}(x) & = & d_{N,i}^{(N_f,\alpha)}({\bf m})  
P_N^{(N_f-1,\alpha)}(-m_i^2;{\bf m}_{\neq i})
P_N^{(N_f-1,\alpha)}(x;{\bf m}_{\neq i}) \nonumber \\
 & = & k_{N}^{(N_f,\alpha)}({\bf m})
\frac{P_N^{(N_f-1,\alpha)}(x;{\bf m}_{\neq i})}
{k_{N}^{(N_f-1,\alpha)}({\bf m}_{\neq i})} 
\frac{\sqrt{w^{(N_f,\alpha)}(x)}}{(x+m_i^2)} \ , \label{finalpsii}
\eeqn
for $i=1,\ldots,N_f$ and
\beq
\label{finalpsi0}
\psi_{2N_f+2}(x) = 
k_{N}^{(N_f,\alpha)}({\bf m})
\frac{P_N^{(N_f,\alpha-1)}(x;{\bf m})} 
{k_{N}^{(N_f,\alpha-1)}({\bf m})}
\frac{\sqrt{w^{(N_f,\alpha)}(x)}}{x} \ ,
\eeq
for $i=N_f+1$. \\
Finally, let us remark that such compact forms 
eq.~(\ref{finalpsii}) and eq.~(\ref{finalpsi0})  are 
quite useful and effective in calculating the microscopic limit of 
$\psi$ functions.

\end{appendix}

\end{document}